\documentclass{emulateapj} 

\usepackage{graphicx}
\usepackage{psfig}

\received{} 
\revised{}
\accepted{}

\shorttitle{WHIM Emission} 
\shortauthors{Fang et al.}

\begin{document}

\title{Simulation of Soft X-ray Emission Lines from the Missing
Baryons}

\author{Taotao~Fang\altaffilmark{1,8},
  Rupert~A.~C.~Croft\altaffilmark{1},
  Wilton~T.~Sanders\altaffilmark{2}, John~Houck\altaffilmark{3},
  Romeel Dav\'{e}\altaffilmark{4},  Neal Katz\altaffilmark{5},
  David~H.~Weinberg\altaffilmark{6}, and Lars
  Hernquist\altaffilmark{7}} \altaffiltext{1}{Carnegie Mellon
  University, Department of Physics, 5000 Forbes Avenue, Pittsburgh,
  PA 15213. Current email address: fangt@astro.berkeley.edu} \altaffiltext{2}{University of Wisconsin,
  Madison, Madison, WI, 53706}  \altaffiltext{3}{Center for Space
  Research, Massachusetts Institute of Technology, 70 Vassar Street,
  Cambridge, MA 02139} \altaffiltext{4}{Steward Observatory,
  University of Arizona, Tuscon, AZ 85721} \altaffiltext{5}{Department
  of Physics and Astronomy, University of Massachusetts, Amherst, MA
  01003} \altaffiltext{6}{Department of Astronomy, The Ohio State
  University, Columbus, OH 43210} \altaffiltext{7}{Harvard-Smithsonian
  Center for Astrophysics, 60 Garden Street, Cambridge, MA 02138} \altaffiltext{8}{{\it Chandra Fellow}}

\begin{abstract}
We study the soft X-ray emission (0.1 -- 1 keV) from the Warm-Hot
Intergalactic Medium (WHIM) in a hydrodynamic simulation of  a Cold
Dark Matter universe. Our main goal is to investigate how such
emission can be explored with a combination of imaging and
spectroscopy, and to motivate future X-ray missions.  We first present
high resolution images of the X-ray  emission in several energy bands,
in which emission from different ion species dominates.  We pick three
different areas  to study the high resolution  spectra of X-rays from
the warm-hot IGM: (a) a galaxy group; (b) a filament and (c) an
underluminous region. By taking into account the background X-ray
emission from AGNs and foreground emission from the Galaxy, we
compute composite X-ray spectra of the selected regions. We  briefly
investigate angular clustering of the soft-X-ray emission, finding a
strong signal. Most interestingly, the combination of high spectral
resolution and angular information allows us to map the emission from
the WHIM in 3 dimensions. We cross-correlate the positions of galaxies
in the simulation with this redshift map of  emission and detect the
presence of six different ion species (Ne IX, Fe XVII, O VII, O VIII,
N VII, C VI) in  the large-scale structure traced by the
galaxies. Finally we show how such emission can be detected and
studied with future X-ray satellites, with particular attention to a
proposed mission, the Missing Baryon Explorer, or $\sl
MBE$. We present simulated observations of the WHIM gas with $\sl MBE$.
\end{abstract}


\section{Introduction}

Cosmological hydrodynamic simulations provide us with  quantitative
predictions for the state of baryonic matter in the universe (see,
e.g, \citealp{cos99,dav01}). At high redshift ($z\gtrsim 3$), most
baryons are diffuse, with a density close to the cosmic mean, and are
located in the intergalactic medium (IGM).  They can be probed readily
using Ly$\alpha$ absorption systems in the optical spectra of high
redshift quasars (see e.g.,  \citealp{cen94,zan95,hkw96,wmhk97}). At
low redshift, the process of gravitational  collapse has caused a
minority of baryons to condense into structures such as stars,
galaxies, groups and clusters of galaxies. While a  substantial
fraction ($\sim$ 30\%) of baryons are trapped within the gravitational
potential wells of these large scale structures in the form of hot
intracluster or intragroup gas, the remaining baryons are left in
intergalactic space and are predicted to form filamentary structures
seen in cosmological hydrodynamic simulations. Some of these
structures contain baryons with relatively low temperatures ($T
\lesssim 10^4$ K), and therefore enough residual neutral hydrogen to
be detectable as the low-redshift Ly$\alpha$ forest (see, e.g.,
\citealp{dhkw99,pss00}). However,  a larger fraction have been
shock-heated to temperatures between $10^5 - 10^7$ K. Given these
temperatures, these baryons, which form the so-called ``warm-hot
intergalactic medium,'' or ``WHIM,'' are best probed using UV/X-ray
observations.

Recently, important progress has been made in the detection of
absorption features produced by highly ionized metals in this WHIM gas
in the UV/X-ray spectra of background quasars. In the UV band, a
significant number of \ion{O}{6} absorption lines have been seen with
the Far Ultraviolet Spectroscopic Explorer ($\sl FUSE$) and the Hubble
Space Telescope ($\sl HST$) (see, e.g.,
\citealp{stl98,tsj00,tsa00,ssr02}). The distribution and derived
properties of these \ion{O}{6} lines are consistent with predictions
from simulations (see, e.g., \citealp{fbr01,cto01,che02}). In the
X-ray band, \citet{fan02}, \citet{mwc02}, \citet{cag02} and \citet{mck03} reported
on the detection of intervening \ion{O}{7} and/or \ion{O}{8}
absorption lines with {\sl Chandra} and {\sl
XMM}-Newton. \citet{nic02}, \citet{fsc03} and \citet{ras03} also
reported the detection of $z \approx 0$ X-ray absorption lines. These
low redshift lines may be attributable, at least in part, to the WHIM gas in
our Local Group.

The absorption line method does have the advantage of directly probing
the absorbing ions, as the detected line strength is proportional to
the column density and so to the ion number density. However, it
suffers the limitation of being only able to probe one dimensional
information, along limited lines of sight, and detectability is also
constrained by the flux of background sources. In order to fully
reveal the three dimensional structure and physical properties of the
WHIM gas, we need to study its {\it emission} with
imaging/spectroscopic methods. In this paper, we will focus on the
X-ray emission from the WHIM gas. In general, X-ray emission from a
hot, diffuse plasma contains two parts, a continuum which is contributed
by various emission mechanisms, and line emission from metals
in the IGM. While we will briefly discuss the continuum emission from
the WHIM, this paper will be devoted mostly to studying the line
emission.

Several authors have extensively studied the broad band soft X-ray
emission (0.5 -- 2 keV) from the diffuse background, particularly the
X-ray emission from the WHIM gas, using hydrodynamical simulations
(see, e.g., \citealp{cko95,poc01,cro01}). Much effort has been put
into investigating the overall intensity of the emission, and how to
distinguish it from the two dominant sources of the soft X-ray
background: X-ray emission from  active galactic nuclei (AGNs) and
from the hot gas in our Galaxy (see, e.g,
\citealp{ksn00,ksm01}). Simulations predict that the mean intensity of
emission from the WHIM gas is between (2 -- 4) $\times 10^{-13}\rm\
ergs\ s^{-1}cm^{-2}deg^{-2}$, only about 5 -- 15\% of the total
extragalactic X-ray emission between 0.5 -- 2 keV
(\citealp{poc01,cro01}). While it seems to be a difficult task to
detect most of this WHIM emission with current X-ray telescopes,  it
appears feasible  to  detect the high intensity tail of the WHIM X-ray
intensity distribution.  For instance, \citet{sch00} reported on the
presence of a possible filament with intensity $\approx 6 \times
10^{-13}\rm\ ergs\ s^{-1}cm^{-2}deg^{-2}$ with the {\sl ROSAT} PSPC;
and \citet{mar03} and \citet{kaa03} reported detecting \ion{O}{7}
and/or \ion{O}{8} from an extended, diffuse structure, possibly from
WHIM gas; and \citet{mcc02} recently performed a large-field ($\sim$ 1 sr),
high spectral resolution (5 -- 12 eV) observation with a  sounding
rocket and detected \ion{C}{5}/\ion{O}{7}/\ion{O}{8}  lines from the
diffuse background, some fraction of which may be from the WHIM at
zero redshift.

We expect that future X-ray missions will enable high resolution imaging/spectroscopic observations of the emission from the WHIM gas. We explore such a scenario with a high resolution hydrodynamic
simulation in this paper. Our purpose is to address two important questions: with high spectral/spatial resolution, what does the X-ray emission from the diffuse, hot gas in intergalactic space look like; and are proposed future X-ray missions capable of conducting useful observations? Particularly, we will focus on a proposed mission called the Missing Baryon Explorer, or $\sl MBE$. Given its large field-of-view ($29.5\arcmin\times29.5\arcmin$), high spectral ($\sim 4$ eV) and moderate angular ($\sim
5\arcmin\times5\arcmin$) resolution, we find that $\sl MBE$ will be
capable of revealing the X-ray emission from the WHIM gas. In
addition, we will explore the clustering properties of the WHIM gas,
and its correlation with nearby large scale structures traced by
galaxies.  

Recently, \citet{yos03} examined the detectability of the X-ray emission from the WHIM gas independently with a similar numerical approach. They presented a detailed study of \ion{O}{7} and \ion{O}{8} emission lines from WHIM. They also investigated a dedicated X-ray mission, {\sl DIOS} (Diffuse Intergalactic Oxygen Surveyor). It turns out that {\sl DIOS} and {\sl MBE} very similar properties.  While there are some differences between our work and theirs, such as the selection of the energy bands (\citealp{yos03} concentrated on 0.5 -- 0.7 keV while we look at a broader band between 0.1 -- 1 keV) and the choice of the IGM metallicity models, where they can be compared our results agree with each other reasonably well.

Our plan for this paper is as follows.  We give a brief explanation of
the simulation method and our treatment of the simulation data in
\S\ref{simu}. A full study of the emission from the WHIM gas is
presented in \S\ref{soft}, followed by measurement of its auto- and
cross-correlation properties, making use of the galaxy distribution in
the simulation for the latter.  We investigate the detectability of
emission with the proposed {\sl MBE} X-ray telescope in \S\ref{dete},
and \S\ref{disc} is our discussion and summary.

\section{Simulation}\label{simu}

We use a cosmological hydrodynamic simulation studied by \citet{dav01}
(their model D1). It was run with PTreeSPH, or parallel tree smoothed
particle hydrodynamics (see \citealp{ddh97}). We refer the reader to
these two papers as well as \citet{cro01} for more details.  The model
simulated is a Cold Dark Matter (CDM) universe with a cosmological
constant ($\Lambda$CDM), with cosmological parameters
$\Omega_{\Lambda}=0.6$ and $\Omega_{m}=0.4$. The parameter $h=0.65$,
where the Hubble constant is $H_{0}=100h\rm\ km\
s^{-1}Mpc^{-1}$. The simulation box size is  $50h^{-1}\rm Mpc$ with a
$7h^{-1}\rm kpc$ spatial resolution (comoving units, equivalent
Plummer softening).  The baryonic mass resolution is $8.5 \times 10^8\
M_{\odot}$, where $M_{\odot}$ is the solar mass.  Cooling from H and
He was included, as well as star formation and mild feedback. As
discussed in \citet{dav01}, the feedback consists of thermal energy
which is released into star forming regions, which are of high
density. The energy is rapidly radiated away and has little effect on
the IGM.  The simulation was run from redshift  $z=49$ to the
present. We make use of 27 of the outputs,  roughly logarithmically
spaced in redshift,ranging from $z=5.98$ to $z=0$.

For each gas particle, we calculate the X-ray emissivity based on a
Raymond-Smith code \citep{rsm77}. Note that we have first recalculated
the SPH densities for each X-ray emitting particle after excluding the
cold particles from the density  estimator (see \citealp{cro01} for
details).  This decoupling of hot and cold phases (see also
\citealp{pearce}, for example)  is necessary in order avoid
overestimating the  densities of hot particles which are close to cold
particles (see also \citealp{she02}).

\begin{figure}
\includegraphics[angle=90,scale=0.37]{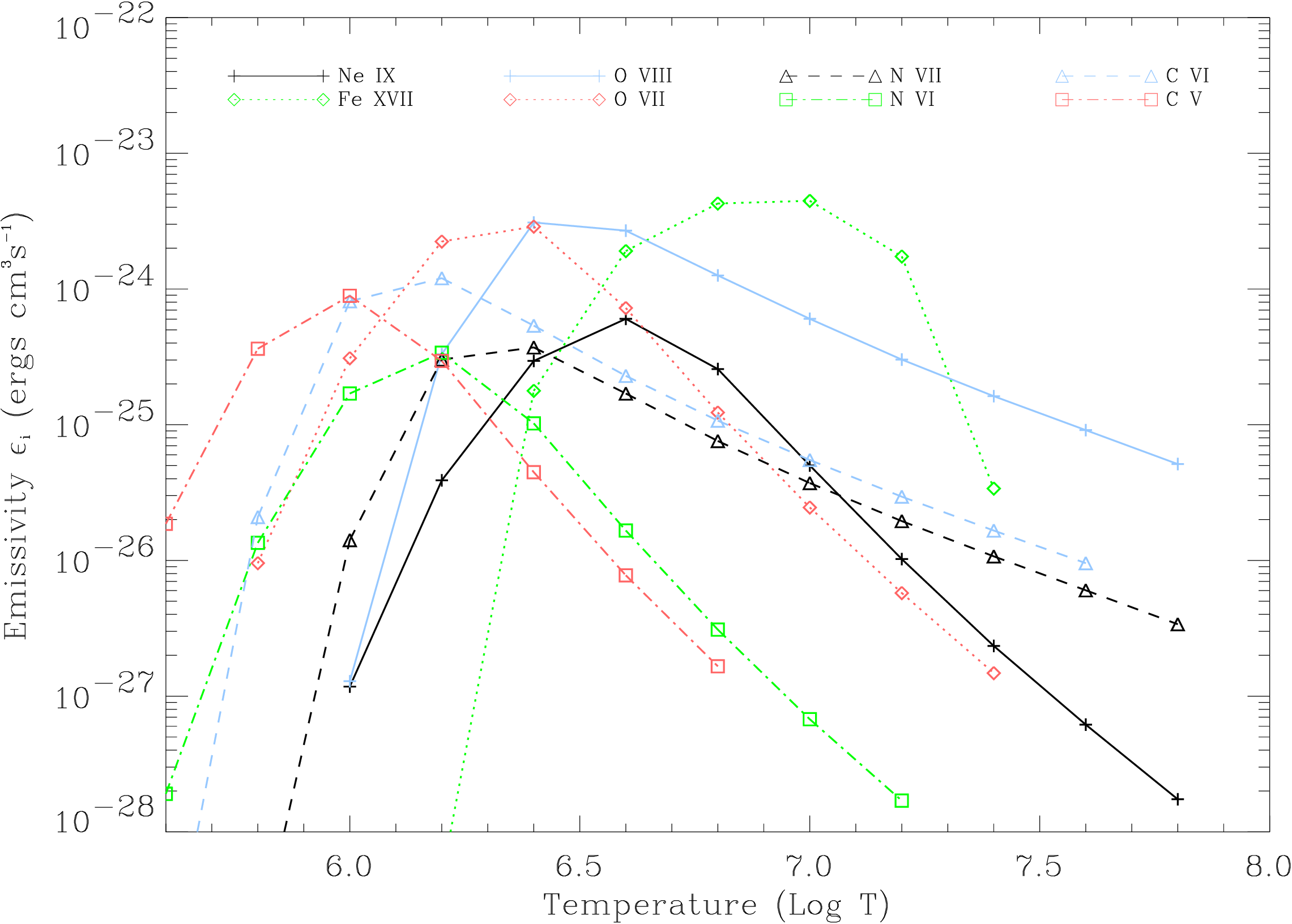}
\caption[h]{X-ray emissivity for eight ion species. Among them,
\ion{O}{7}, \ion{O}{8}, and \ion{Fe}{17} show the highest peak
emissivities between $10^6$ -- $10^7$ K. Emission models were taken from
\citet{rsm77}.}
\label{f1}
\end{figure}

 We adopt a density-dependent metallicity, $Z \propto
\sqrt{\rho/\overline{\rho}}$, to mimic the density-metallicity
relation obtained by \citet{cos99},  \cite{she03}, and Aguirre et
al. (2001a,b,c) in their simulations. Here $\rho$ and
$\overline{\rho}$ are the density and the mean density of the
universe. We normalize this relation by setting $Z = 0.005 Z_{\odot}$
at $\rho = \overline{\rho}$ to match  Ly$\alpha$ forest measurements,
and limiting the metallicity to be  $Z \leq 0.3 Z_{\odot}$ in  much
more overdense regions to match the observational results from galaxy
clusters in the local universe. Here $Z_{\odot}$ is the solar
abundance. The Raymond-Smith code includes eleven heavy elements,
namely C, N, O, Ne, Mg, Si, S, Ar, Ca, Fe, and Ni. Table~1 lists the
wavelengths and the corresponding energies of the strongest emission
lines we are probing with our simulation. Assuming collisional
ionization equilibrium, Figure~\ref{f1} shows the emissivity
$\epsilon_i$ for each ion species \citep{rsm77}, where $i$ represents
ion species $X_i$. We give the strongest line (the resonance line in
the He-like triplet), and for \ion{Fe}{17} we select the 825.79 eV
line.

\begin{figure*}
\includegraphics[angle=90,width=1.0\textwidth,height=0.4\textheight]{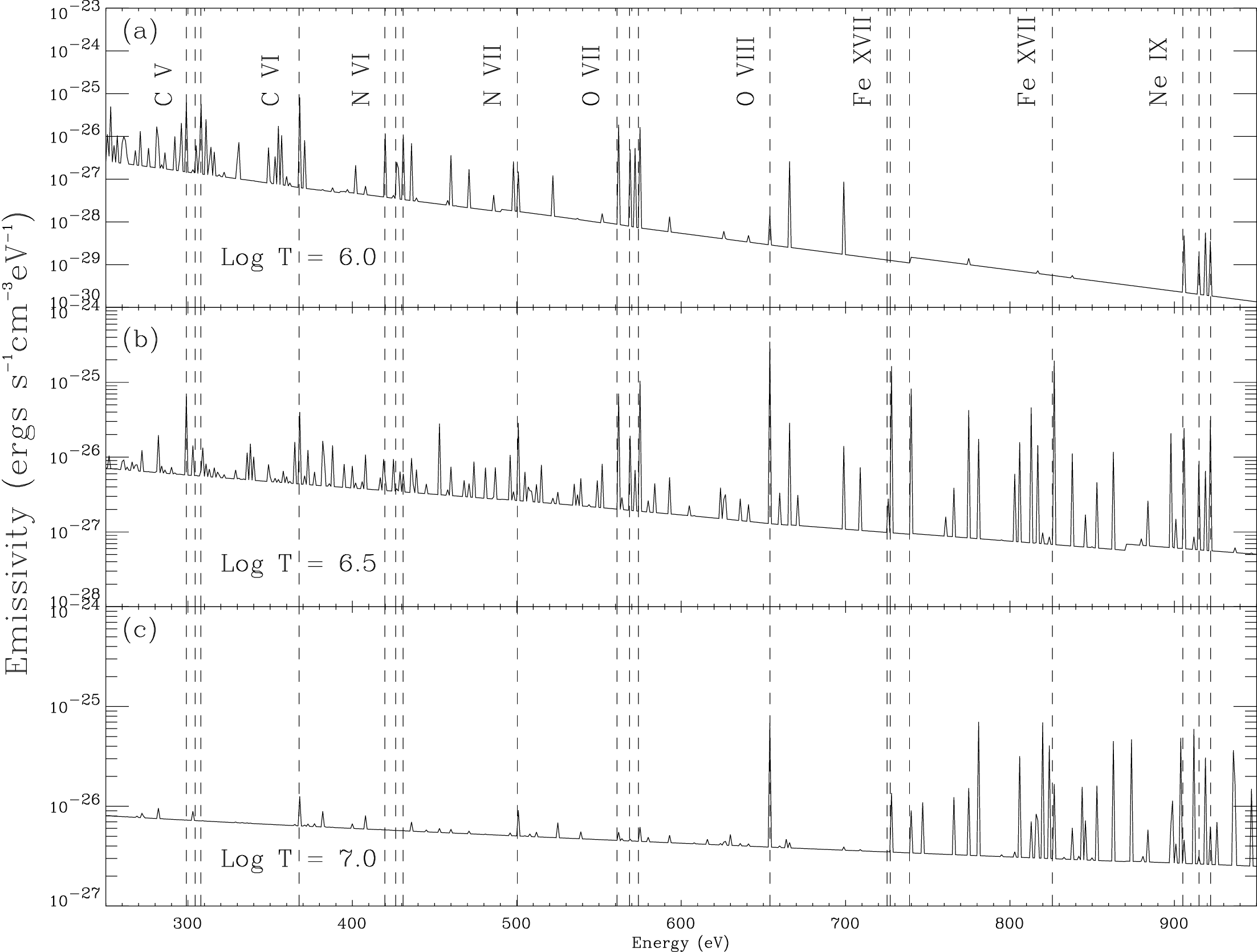}
\caption[h]{X-ray emissivity for hot gas at three  temperatures
typical of the $z=0$ IGM: (a) T = $10^6$ K; (b) T = $10^{6.5}$ K; (c)
T = $10^7$ K. The gas has a metallicity of 0.1 $Z_{\odot}$. Emission
models were taken from \citet{rsm77}.}
\label{f2}
\end{figure*}

In Figure~\ref{f2} we display three template  X-ray emission spectra
for hot gas at different temperatures: (a) T = $10^6$ K; (b) T =
$10^{6.5}$ K; (c) T = $10^7$ K. We use a metallicity of 0.1
$Z_{\odot}$, which is typical for gas with overdensities $\delta \sim
$ a few hundred based on simulations (\cite{cos99},
\cite{she03}). Here $\delta$ is defined as $\delta \equiv
\rho/\overline{\rho}-1$. Figure~\ref{f2}a and ~\ref{f2}b show that
at low temperatures the spectra are dominated by C, N and \ion{O}{7}
emission lines. At high temperatures \ion{O}{8} and Fe emission start
to take over. At temperatures above $10^7$ K, the spectra are
completely dominated by Fe and Ne emission lines.

\vbox{
\begin{center}

\begin{tabular}{lcccc}
\multicolumn{5}{c}{~~~~~~~~Table 1: Emission Line List~~~~~~} \\
\hline \hline Ion & Wavelength ($\AA$) & Energy (eV) & $\rm Line^1$ &
$\rm Reference^2$ \\  \hline \ion{C}{5} & 41.47 & 298.96 & f & a\\  &
40.73 & 304.41 & i & a\\ & 40.27 & 307.89 & r & a\\ \ion{C}{6} & 33.73
& 367.54 &   & b\\ \ion{N}{6} & 29.53 & 419.79 & f & a\\ & 29.08 &
426.30 & i & a\\ & 28.78 & 430.80 & r & a\\ \ion{N}{7} & 24.78 &
500.35 &   & b\\ \ion{O}{7} & 22.10 & 560.98 & f & a\\ & 21.80 &
568.62 & i & a\\ & 21.60 & 573.95 & r & a\\ \ion{O}{8} & 18.97 &
654.00 &   & b\\ \ion{Fe}{17} & 17.10 & 725.22 & & c\\ & 17.05 &
727.14 & & c\\ & 16.78 & 738.88 & & c\\ & 15.01 & 825.79 & & c\\
\ion{Ne}{9}& 13.70 & 905.10 & f & a\\  & 13.55 & 915.03 & i & a\\ &
13.45 & 922.00 & r & a\\ \hline
\label{t1}
\end{tabular}

\parbox{5in}{
\vspace{0.1in} \small\baselineskip 9pt
\indent 1. Line: r --- resonance; i --- intercombination; f ---
forbidden\\ 2. Reference: a. \citet{dra88}; b. \citet{jso85};
c. \citet{bro98}\\  }
\end{center}
\normalsize \centerline{} }

Three types of continuum emission make contributions to the total
X-ray continuum we discuss here: thermal bremsstrahlung radiation,
or free-free emission; radiative recombination radiation, or
free-bound emission; and two-photon decay (see. e.g.,
\citealp{rsm77}). Their relative contributions vary with temperature
and photon energy. For instance, at T $\sim 10^6$ K radiative
recombination dominates at E $ > 0.4 $ keV, while thermal
bremsstrahlung radiation dominates over the entire X-ray band when T
$\sim 10^7$ K. The total continuum emissivity $\epsilon_c$, as a
function of photon energy $E$, can be written as \citep{mgr81}:

\begin{equation}
\epsilon_c (E) \propto \overline{G_c}(E,T) E T^{-1/2} n^2_e \exp
(-E/kT),
\end{equation} 
where $T$ is the gas temperature, $n_e$ is the electron density, $k$
is Boltzmann's constant, and $\overline{G_c}(E,T)$ is the total
effective Gaunt factor, which includes all three continuum emission
processes \citep{grm78}. By integrating $ \epsilon_c (E)$ over energy
$E$, the integrated X-ray continuum emissivity is $\propto
n^2_e T^{\alpha}$, where $\alpha=1/2$ for high temperature gas, in
which bremsstrahlung radiation dominates, and  $\alpha=-3/2$ for gas
with $T \lesssim$ a few $\times 10^6$ K.

\section{Soft X-ray Spectra of the Warm-Hot Intergalactic Medium}\label{soft}

\subsection{The IGM: simulated maps and spectra}

In order to go from the three dimensional particle distributions in
the simulation outputs to sky maps and spectra, we rely on the
technique of stacking simulation boxes along the lightcone (see e.g.,
\citealp{dasilva,swh01,cro01}). We place simulation boxes one behind
the other until we reach the comoving  distance corresponding to a
given redshift ($z=1$ in the first case).  The output file
corresponding to the closest redshift is used at  each point (we use
11 different outputs between $z=0$ and $z=1$).  The boxes are randomly
rotated, reflected and translated \citep{cro01} so that periodic
structure does not repeat along the line of sight.  Because of this,
only structure on scales smaller than the box is preserved, and we
only consider structure on these scales when measuring clustering
statistics. For example, the median redshift of emission is $z=0.45$
in the $0.5-2$ keV band, so that we cannot accurately simulate
structure with angular scales $\gtrsim$ 100 arcmins.

The X-ray emissivity in a range of 2eV wide bins in  energy is
computed for each particle, and these quantities are allocated to a
grid using the projection of the SPH kernel on the plane of the sky
(see below for more details). The grid is a three dimensional
datacube, with two angular axes and one axis for photon energy.

\begin{figure*}
\includegraphics[angle=270,scale=0.8]{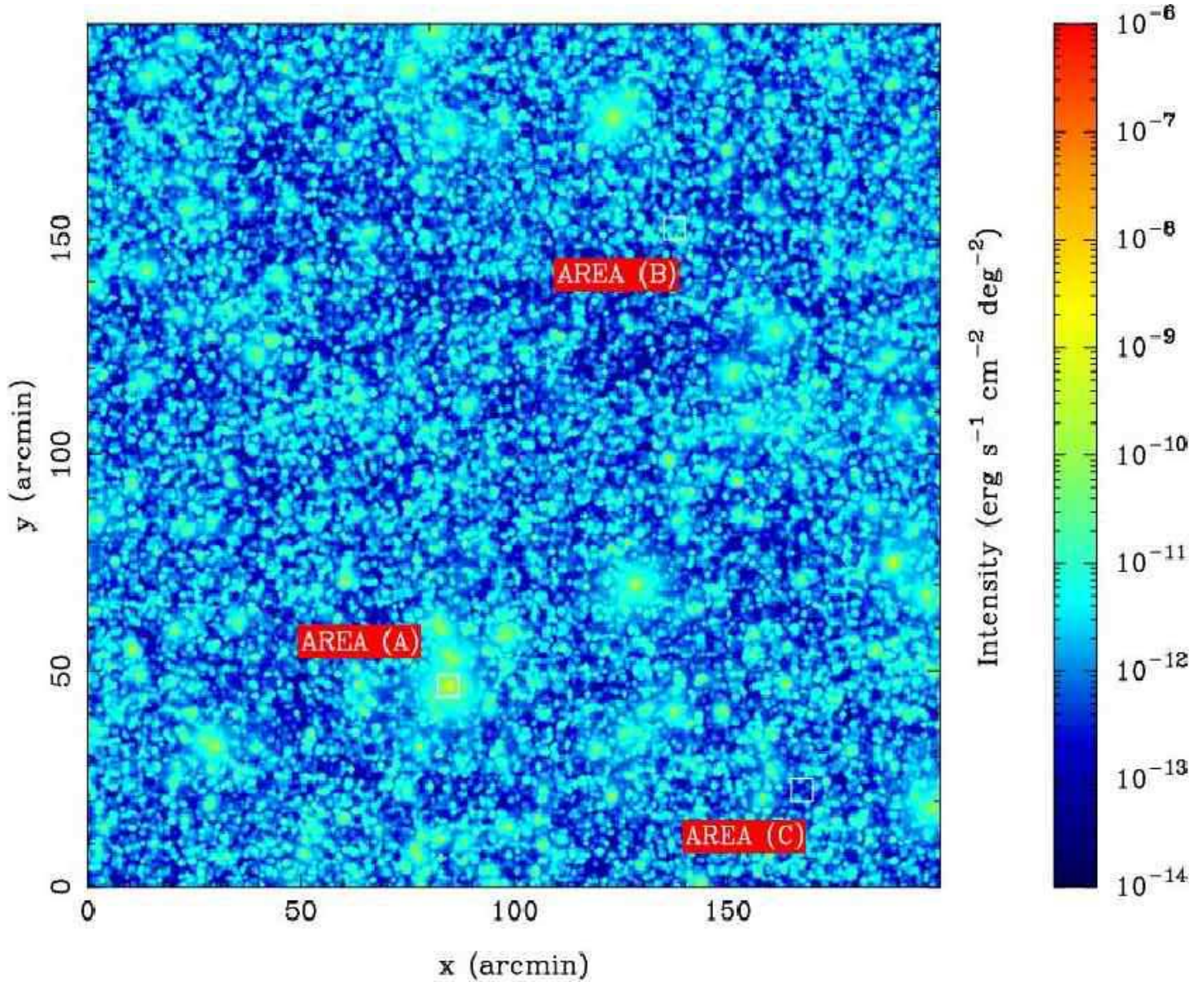}
\caption[h]{The simulated $200\arcmin \times 200\arcmin$ region. The
whole region is divided into 512$\times$512 grid cells, and the X-ray
intensity is each cell is obtained by integrating the X-ray flux
between 0.1 -- 1 keV in that cell. We select three regions (shown as
white boxes) to study three typical types of X-ray emission: (a) hot
gas in a galaxy group; (b) diffuse X-ray emission in a filament; and
(c) a void. \label{f3}}
\end{figure*}

We first examine the angular distribution of the emission in a
relatively broad spectral band.  In Figure~\ref{f3} we show a map
of a $200\arcmin\times200\arcmin$ simulated region. The area has been
divided into a 512$\times$512 grid, so that each pixel has side length
$\sim 23\arcsec$. The X-ray emission includes both continuum and metal
line emission from the hot IGM, integrated between 0.1 -- 1
keV. \citet{cro01} showed  that the majority of the X-ray emission
from the IGM received at $z=0$ comes from gas with $z<1.0$ (about
90\%, see  Figure~11a of \citealp{cro01}). Due to dilution by
projection, filaments that would be apparent in three dimensions do
not appear readily here. The intensity varies from $10^{-6} -
10^{-7}\rm\ erg\ s^{-1}cm^{-2}deg^{-2}$, in patches of sky dominated
by emission from the hot IGM in groups or clusters of galaxies, down
to $10^{-14}\rm\ erg\ s^{-1}cm^{-2}deg^{-2}$, in regions of low
density. We label three ($5\arcmin\times5\arcmin$) areas that we will
study in detail spectroscopically in the following sections: area (a),
with an average surface brightness of $10^{-7}\rm\ erg\
s^{-1}cm^{-2}deg^{-2}$, representing X-ray emission from a galaxy
group; area (b), with a mean surface brightness of $10^{-9}\rm\ erg\
s^{-1}cm^{-2}deg^{-2}$, representing X-ray emission from a
superposition of  filamentary structures; and area (c), with a surface
brightness of $10^{-13}\rm\ erg\ s^{-1}cm^{-2}deg^{-2}$, an underdense
and void-like region.

\begin{figure*}
\includegraphics[width=1.\textwidth,height=0.65\textheight]{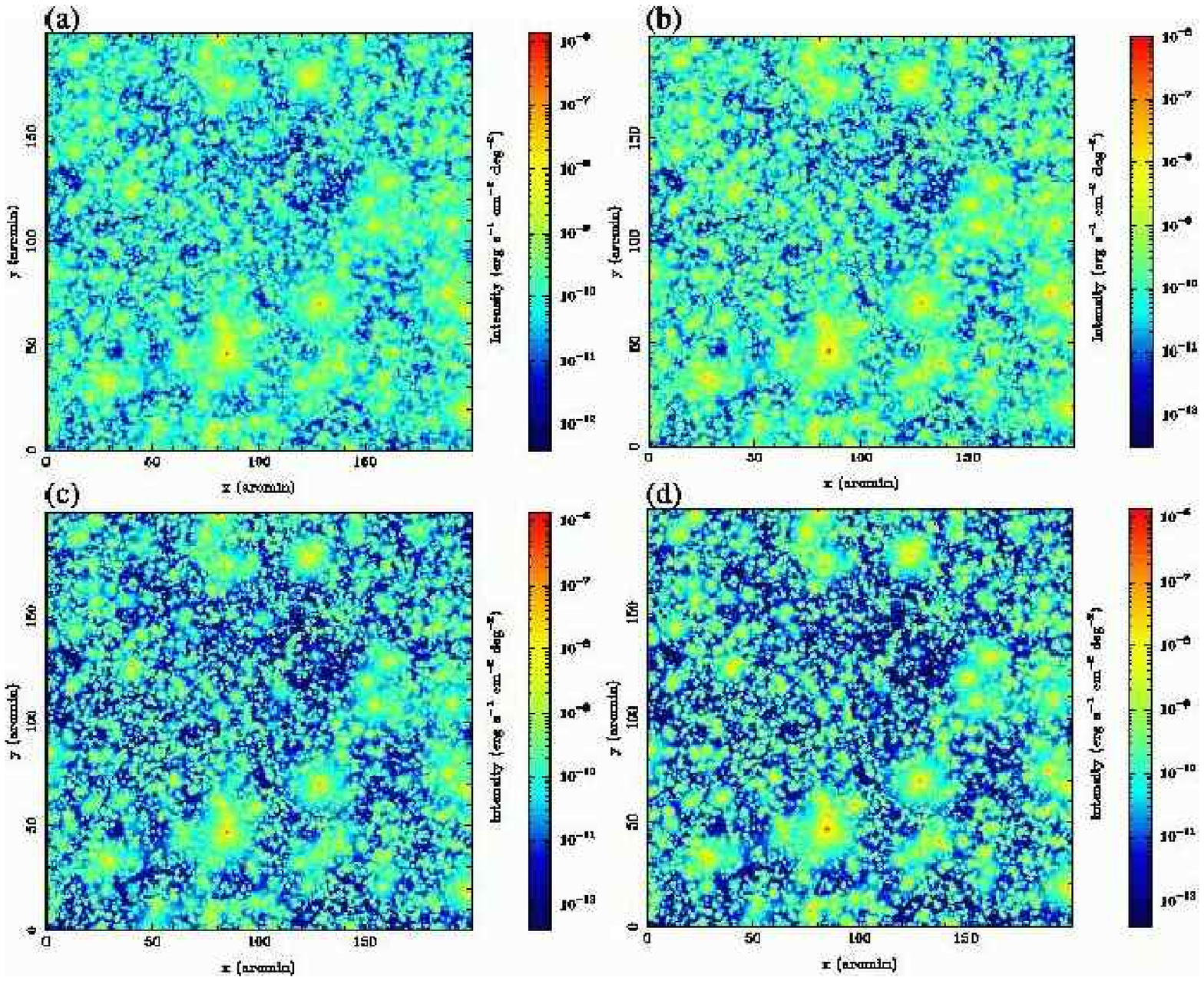}
\caption[h]{The simulated $200\arcmin \times 200\arcmin$ region in
four energy bands: (a) 250 -- 350 eV; (b) 350 -- 425 eV; (c) 550 --
600 eV; (d) 700 -- 800 eV. Emission lines from C, N, O, and Fe
dominate in each band, respectively.}
\label{f4}
\end{figure*}

Figure~\ref{f4} shows the soft X-ray emission from the same sky
area but in four narrower energy bands. Since the X-ray continuum
emissivity exponentially decreases with energy, in most energy intervals the
flux from the X-ray emission lines dominates over the continuum
emission. In panel (a) emission lines from carbon, particularly
\ion{C}{6} and \ion{C}{5} triplet dominate from 250 to 350 eV. Here
we assume that most X-ray emission lines come from the low-redshift
IGM. \citet{cro01} showed that X-ray emission from the IGM peaks at $z
\sim 0.2$ and decreases slowly to high redshift. Panel (b) shows the
band (350 -- 425 eV) in which nitrogen lines dominate. Panel (c) is
for the band (550 -- 600 eV) for which oxygen lines dominate (mainly
\ion{O}{7} triplet). Panel (d), the highest energy is in the band
where emission from iron is very strong, mainly the Fe L complex. It
can be clearly seen from the emission maps in these four narrower
energy bands that low energy emission is rather more diffuse
(panels a and b) than high energy emission, as would be expected from  a
superposition of filamentary structures. On the other hand, at high
energy emission comes from more obviously discrete sources, such as
collapsed regions which form groups of galaxies.

To study the spectroscopic properties of the soft X-ray emission from
the WHIM gas, we divide the simulated patch of sky into $40\times40$
grid cells, with each cell extending over an area of
$5\arcmin\times5\arcmin$
\footnote{The pixel area of $5\arcmin\times5\arcmin$ is the spatial
resolution of the Missing Baryon Explorer ({\sl MBE}), a proposed
X-ray mission which we will discuss later.}. In every cell, we obtain
the X-ray spectrum of each gas particle by applying the Raymond-Smith
code, based on its temperature, density, metallicity and redshift. The
peculiar velocities of the gas particles are used in order to produce
a redshift space spectrum.  We then combine all the spectra to obtain
a synthesized spectrum in each cell, as mentioned above.

\begin{figure*}
\includegraphics[width=1.\textwidth,height=0.75\textheight,angle=270]{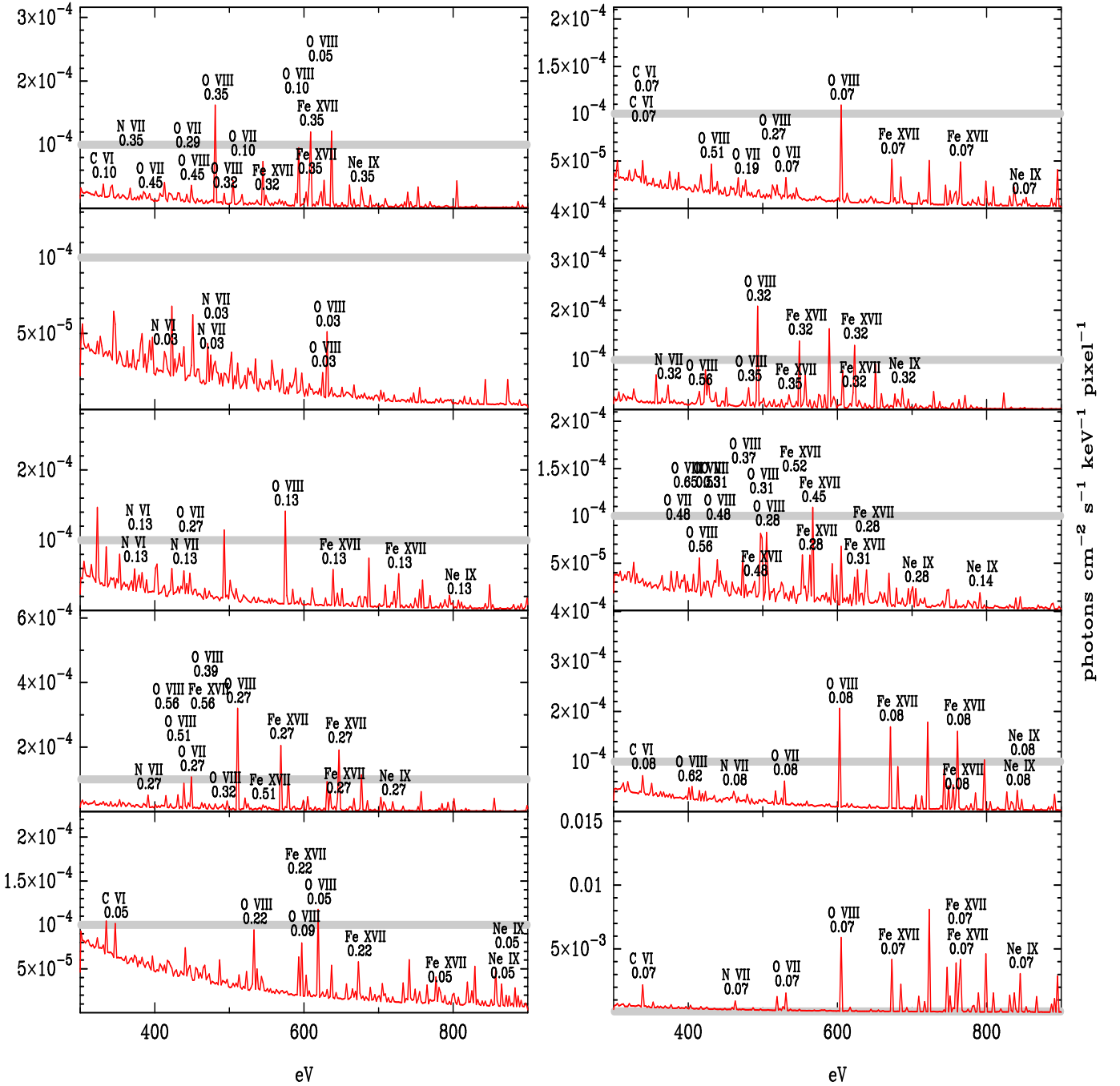}
\caption[h]{10 random sample spectra taken from the brightest 10\% of
pixels.  Note that the y-axis scale changes between each plot. We have
plotted a gray bar at count level $10^{-4}$ photons cm$^{-2}$ s$^{-1}$
pixel$^{-1}$ keV$^{-1}$ so that the panels can more easily be
compared. The pixel size is $5\arcmin\times5\arcmin$. The redshifts of
material contributing to emission lines are shown}
\label{f5}
\end{figure*}

Figure~\ref{f5} shows 10 random samples of such spectra picked from the brightest 10 \% of pixels over the range 0.3--0.9 keV.  A large number of emission lines are visible in many of the panels.  Having information from the underlying simulation allows us to identify lines easily. Our scheme for doing this automatically is as follows. When creating the spectra, for each energy bin, we keep track of the redshift of the particle that was the largest contributor of  X-ray flux to that bin. When plotting the spectra, we divide the energy of bin $i$ by $1+z_{i}$, where $z_{i}$ is the redshift of the maximum flux particle. If the resulting value is within  5 eV  of one of the known ion rest wavelengths  (see Table 1), and the pixel corresponds to a peak in the  spectrum, we label it with the name of ion species, and the redshift $z_{i}$.  It can be seen that many of the features come from the same redshift, but are signposted by
  different species.  For example, the top left panel shows that a
  number of lines (\ion{N}{7}, \ion{O}{8}, \ion{Fe}{17} and
  \ion{Ne}{9}) arise from hot gas at redshift $\sim 0.35$. These types
  of spectra can potentially provide us with  important information
  that could help us identify emission lines  with ion species and
  redshifts in real observations.

\subsection{Contribution from the Galactic Foreground and AGNs} 

{\sl ROSAT} (\citealp{sno95,sno97}) and other X-ray missions (see,
e.g., \citealp{mcc02,san01}) show that the soft X-ray background in
the band 0.1 - 1 keV is comprised of at least three components: (1)
unabsorbed X-ray emission, referred to as the "local hot bubble"
(LHB), from a cavity in the local interstellar medium
(\citealp{sno98,sfe99}); (2) hotter absorbed X-rays, referred to as
"transabsorption emission" (TAE, \citealp{ksn00,sfk00}), from gas in
either the Galactic halo or the intragroup medium of the Local Group
(\citealp{rpe01,ras03}), and (3) an extragalactic X-ray background
(\citealp{has98,mcb00}). Simultaneous fits to ASCA and ROSAT data
arrive at the same components \citep{miy98}.

It is well known that the extragalactic X-ray background is dominated
by discrete sources such as active galactic nuclei (AGNs). {\sl ROSAT}
deep surveys resolved $\sim$70 -- 80\% of the soft X-ray background
(0.5 -- 2 keV) into discrete sources at a flux limit of $10^{-15}\rm\
erg\ cm^{-2}s^{-1}$ \citep{has98}. The {\sl Chandra} deep survey,
which had a flux limit of $3\times 10^{-16}\rm\ erg\ cm^{-2}s^{-1}$,
resolved an additional 6 -- 13\% into point sources
\citep{mcb00}. This leaves room for $\sim$5 -- 25\% to come from
diffuse emission from hot gas in the IGM, which is what we concentrate
on in this paper.

To obtain the composite spectrum of the Galactic foreground and the
AGN background, we use the following simple models of the components.
(1) For the LHB, we adopt a thermal plasma model with k$T = 0.1$ keV
and emission measure $EM = 0.019\rm\ cm^{-6}$ pc. The metal abundance
is taken to be solar, except for Mg, Si, and Fe, which have an
abundance of 0.28 $Z_{\odot}$. (2) For the TAE, we adopt a thermal
plasma model with k$T = 0.227$ keV and emission measure $EM =
0.003\rm\ cm^{-6}$ pc. This model has solar abundances and is absorbed
by a hydrogen column density of $N_H = 1.8 \times 10^{20}\rm\
cm^{-2}$. The parameters for these two thermal components were
obtained from fits to the calorimeter sounding rocket data reported in
\citet{mcc02} using the {\sc MEKAL} spectral models in
XSPEC~\footnote{see http://heasarc.gsfc.nasa.gov/docs/xanadu/xspec/}.
(3) The third component represents the AGN. We used a spectrum
constructed by D. McCammon (private communication) to reflect all the
known ROSAT and Chandra data on the resolved AGN contribution of the
soft X-ray background (see Figure 15, \citealp{mcc02}). In Figure 6 we
show the composite spectrum of these three components.

\begin{figure}
\includegraphics[angle=90,scale=0.35]{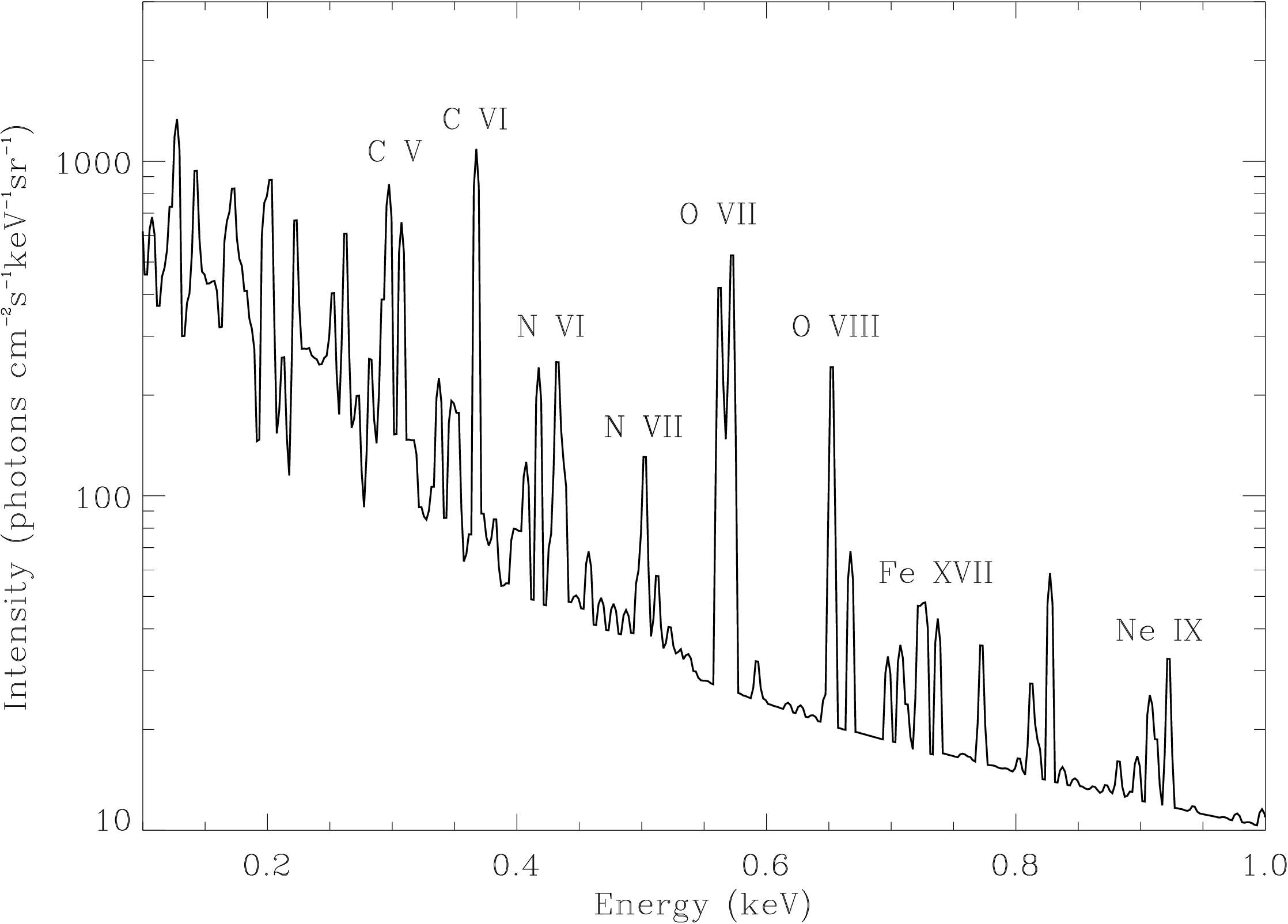}
\caption[h]{Composite spectrum of the Galactic foreground and AGNs. It
includes three components: (a) a thermal emission from the Local Hot
Bubble; (2) an absorbed thermal emission from hot gas either in the
distance halo or in the Local Group; and (c) continuum emission from
discrete AGNs.}
\label{f6}
\end{figure}

\subsection{X-ray spectra of Characteristic Regions}

\begin{figure*}
\includegraphics[angle=90,scale=0.8]{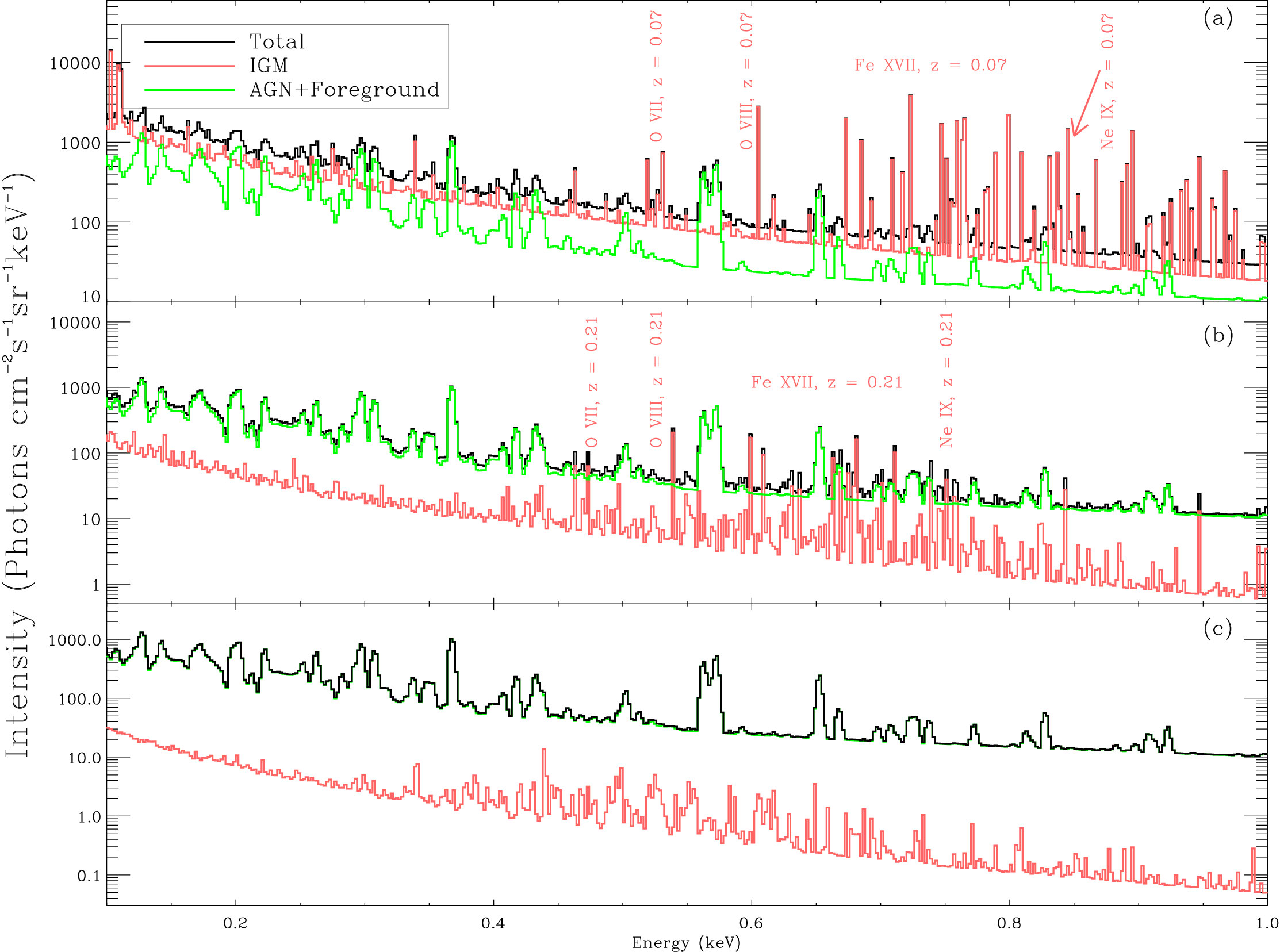}
\caption[h]{Simulated spectra of three selected regions. Green lines
are for the IGM emission, red lines are for the Galactic foreground
plus AGNs, and black lines stand for total X-ray emission. (a) Hot gas
in a galaxy group, it is clear that the IGM emission dominates the
whole spectrum, especially at high energy; (b) diffuse X-ray emission
in a filament; The Galactic foreground plus AGNs' emission dominate,
but some strong emission lines from a filament at $z \approx 0.21$
show up in the total spectrum; and (c) a void, in which the entire
spectrum is dominated by the Galactic foreground plus AGNs. We also
label redshifts of several selected lines.}
\label{f7}
\end{figure*}

We select three regions (shown as white boxes in Figure~\ref{f3})
to study three typical types of X-ray emission: (a) hot gas in a
galaxy group (Figure~\ref{f7}a); (b) diffuse X-ray emission in a
filament (Figure~\ref{f7}b); and (c) a void
(Figure~\ref{f7}c). In each panel we show three curves: the red
line is the emission from the IGM, green is the  composite spectrum of
AGNs and the Galactic foreground, and the black line represents the
total emission. We also identify the most significant IGM emission
lines with their ion species and redshifts in a similar fashion to
Figure~\ref{f5}.

In Figure~\ref{f7}a it is clear that the IGM emission dominates
the entire spectrum. This is not surprising because the X-rays come
from the hot gas within a galaxy group at redshift $z \approx
0.07$. For such hot gas, bremsstrahlung radiation, which dominates
the total X-ray emissivity, is $\propto n_{e}^{2}T^{0.5}$, where $n_e$
is the electron density and $T$ is the temperature. The X-ray emission
from the high density and high temperature intragroup medium shows a
strong continuum, and this dominates over the AGN and Galactic
foreground emission. In this panel we identify emission lines from O,
Fe, and Ne. In particular we can easily make out redshifted \ion{O}{8}
Ly$\alpha$ (resonance line at 611 eV) and the \ion{O}{7} triplet
(forbidden line at 524 eV, intercombination line at 531 eV, and
resonance line at 536 eV). This could provide us with a direct
measurement of the temperature, electron density and  ionization
states of this X-ray emitting gas. The numerous lines between  650 --
850 eV are mainly emission lines from Fe L-shell transitions, and many
of these lines belong to \ion{Fe}{17} transitions.

Figure~\ref{f7}b shows the emission spectrum of a patch of sky
covered with more diffuse structures, known to be filamentary in 3
dimensions [area (b) in Figure~\ref{f3}]. The total spectrum
(black line) is dominated by the background (AGNs) and foreground
emission, especially at E $<0.4$ keV. However, we see several emission
lines from a filament crossing the line of  sight at $z \approx
0.21$. Although the overall continuum level is nearly an order of
magnitude lower than that of the AGNs plus the foreground, these
emission lines are strong enough to poke through the total spectrum
and show up as obvious features.

In Figure~\ref{f7}c the total spectrum is completely dominated
by the AGNs plus the foreground. The emission from the IGM is almost
100 times lower that that of the AGNs plus the foreground, and it is
not possible to detect any signals from the IGM.

\begin{figure}
\includegraphics[angle=270,scale=0.35]{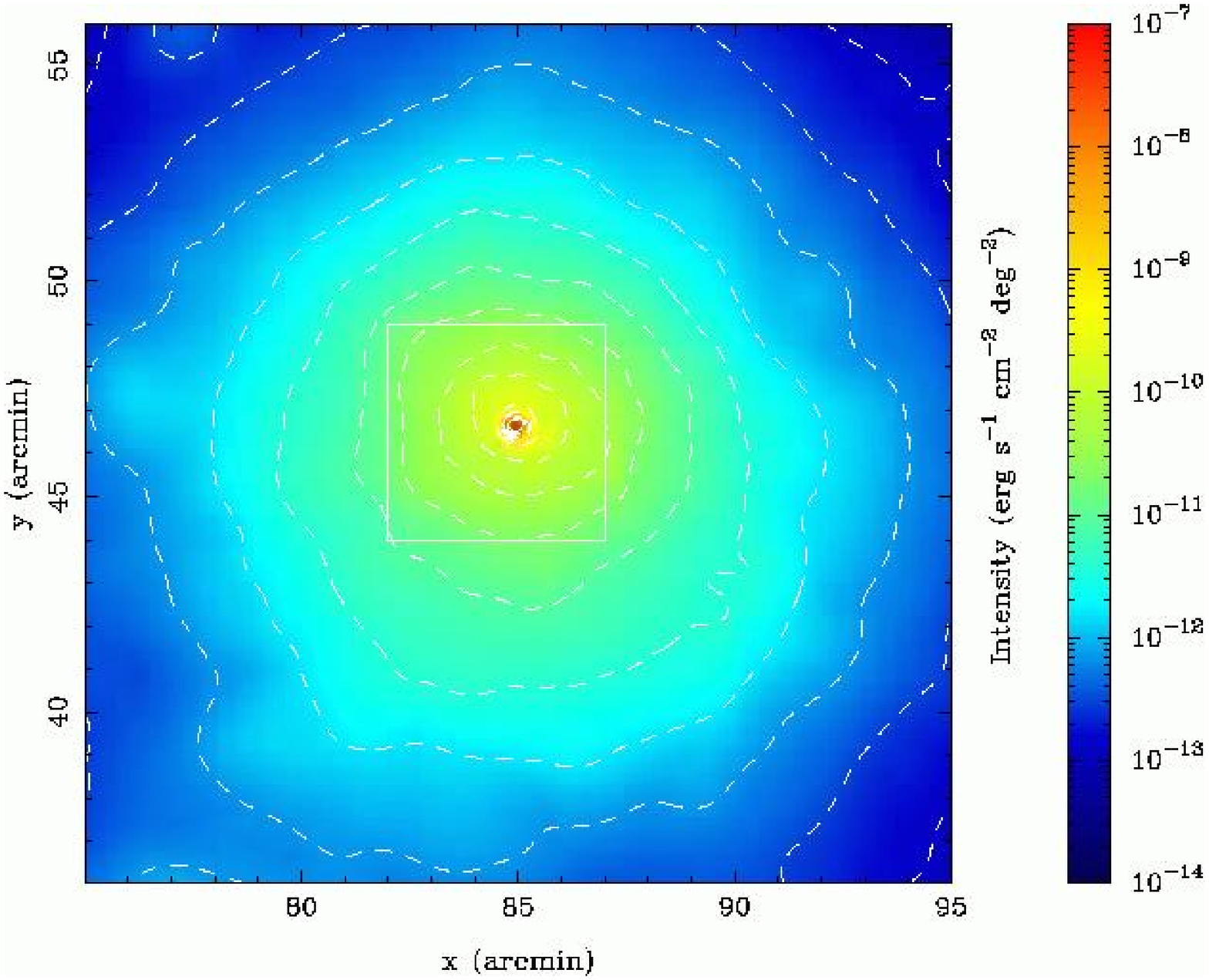}
\caption[h]{A slice of the simulated region, centered on area (a)
(white box), with a size of $20\arcmin\times20\arcmin$. The slice is a
projection of $\Delta z =0.0005$, centered at $z \approx 0.0775$. The
X-ray contours are separated by a factor of $\sim$ 2.2 in intensity,
with the highest contour at a level of $\sim 10^{-7}\rm\ ergs\
s^{-1}cm^{-2}deg^{-2}$.}
\label{f8}
\end{figure}

\begin{figure}
\includegraphics[angle=270,scale=0.35]{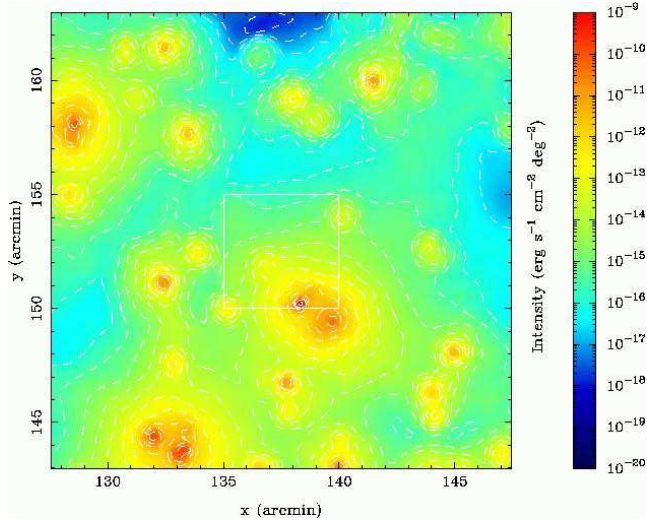}
\caption[h]{Same as Figure~\ref{f8}, but enlarged area (b) (white
box at the center). The slice is a projection of $\Delta z =0.0005$,
centered at $z \approx 0.2108$. The X-ray contours are separated by a
factor of $\sim$ 3.5 in intensity, with the highest contour at a level
of $\sim 10^{-9}\rm\ ergs\ s^{-1}cm^{-2}deg^{-2}$.}
\label{f9}
\end{figure}

In Figure~\ref{f8} and ~\ref{f9} we zoom in on areas (a) and (b),
respectively, plotting only emission from gas which lies within
particular narrow range in redshift (explained below).  These plots
have angular resolution four times higher than that in
Figure~\ref{f3}. Figure~\ref{f8} shows a simulated
$20\arcmin\times20\arcmin$ region, centered on area (a) (denoted by
the white box). Most X-ray emission lines from this area originate
from hot gas at $z \approx 0.0775$, so that we have cut out a slice
that is centered on this redshift with $\Delta z =0.0005$ (about 1.5
comoving $h^{-1}{\rm Mpc}$). Figure~\ref{f8} shows the X-ray
emission from this slice, and the dotted lines show contours of X-ray
emission. In similar fashion we have made Figure~\ref{f9} for area
(b). Clearly, the round shape of the X-ray contours in
Figure~\ref{f8} indicates that the hot gas in area (a) sits in a
virialized, relatively stable system, such as a galaxy group.  The
X-ray contours in area (b) (Figure~\ref{f9})  (we are plotting  only
emission coming from close to $z=0.21$) show rather more diffuse and
irregular X-ray emission from filamentary structures.

\begin{figure}
\includegraphics[angle=90,scale=0.8]{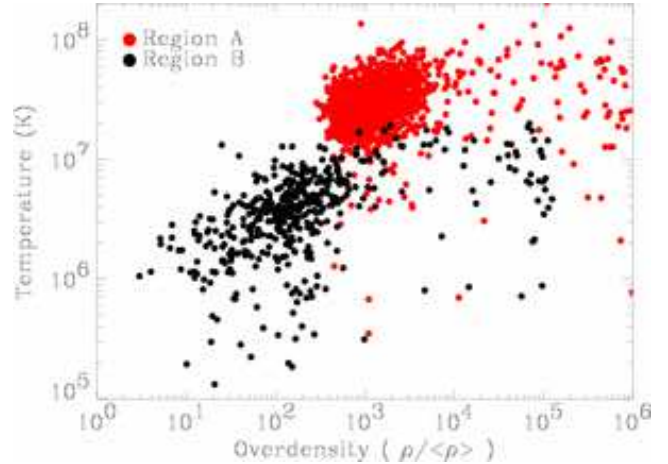}
\caption[h]{Temperature vs. overdensity of gas particles with $T > 10^5$ K in area (a) and (b). Red filled circles represents particles in area (a), and black ones are from area (b).}
\label{f10}
\end{figure}

To further understand the gas distribution within area (a) and (b), we examine all the particles in these two areas centered around redshifts of $\sim$ 0.775 and 0.21, respectively. The length along the line of sight correspond to 5$\arcmin$ at that redshift, i.e., two cubes with comoving sizes $0.33h^{-1}\times0.33h^{-1}\times0.33^{-1}\,\rm Mpc^3$ at $z=0.775$ and $0.86h^{-1}\times0.86h^{-1}\times0.86^{-1}\,\rm Mpc^3$ at $z=0.21$.  In Figure~\ref{f10} we plot the temperature vs. overdensity of gas particles with $T > 10^5$ K. Red filled circles represents particles in area (a), and black ones are from area (b). In area (a), most particles concentrate on a region with overdensities around $10^3$, while in area (b) gas particles distribute more diffusely. The confirms our initial guess that particles in area (a) are more concentrated and tend to form a virialized structure, and in area (b) structure is more filamentary-like.

\begin{figure}
\includegraphics[angle=90,scale=0.35]{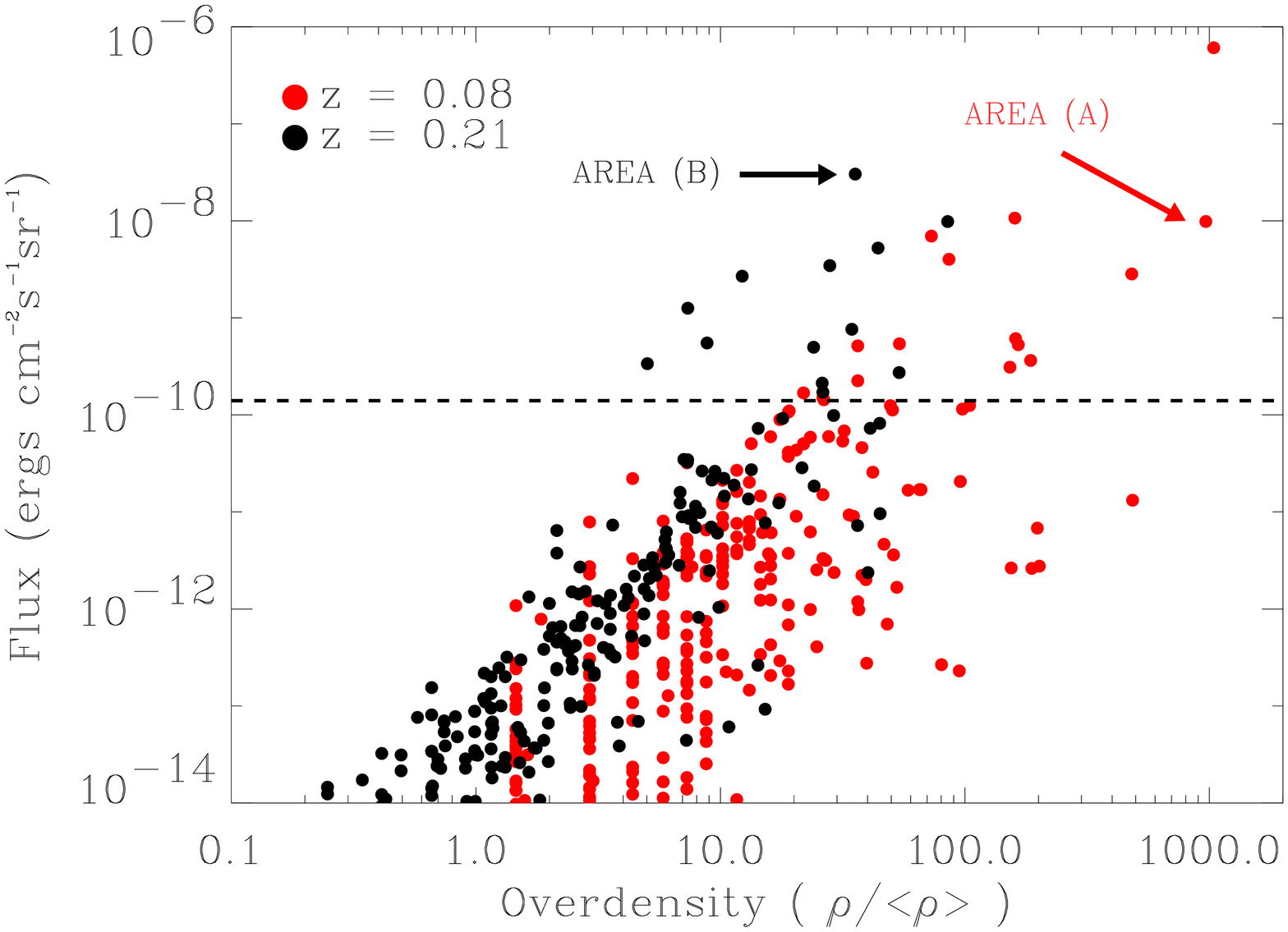}
\caption[h]{\ion{O}{8} line flux vs density in each cube at redshift 0.0775 (red) and 0.21 (black) in our $200\arcmin \times 200\arcmin$ simulation. We also label the two areas (a) and (b). The dashed horizontal line is the estimated flux sensitivity of the proposed X-ray mission {\sl MBE} for a 200 $ksec$ exposure time.}
\label{f11}
\end{figure}

Clearly, an important method to distinguish between area (a) and (b) is to study the emission line strength via, i.e., Figure~\ref{f7}. But exactly how does the observed line strength vary with gas density? To understand this, we select \ion{O}{8} line at 654 eV as a representative line and study the line flux in each of the $5\arcmin \times 5\arcmin$ grid.  Again we select all the grids around the two redshifts, 0.0775 and 0.21, to calculate the \ion{O}{8} flux from each cube. The total flux of \ion{O}{8} emission line from one cube can be obtained by direct summarization of the line flux from each individual particle in that cube \citep{yos03}. The mean density of that cube is calculated by averaging the total mass of particles with $T > 10^5$ K within that cube. Figure~\ref{f11} shows the line flux distribution for both regions: red filled circles represent cubes at redshift of $\sim 0.0775$, and black filled circles are from cubes at $z \sim 0.21$. Two circles pointed by arrows are area (a) and (b). While the line flux dependency on gas density is a complicated relation because it involves properties like temperature, ionization structure and metallicity, in general Figure~\ref{f11} shows that high density regions tend to have high \ion{O}{8} fluxes. We also calculate the flux sensitivity of the proposed X-ray mission {\sl MBE} (the horizontal dashed line) for a 200 ksec exposure time. While only very few girds in this plot are detectable with {\sl MBE} at the two redshifts, the cumulative effect by observing each grid up to higher redshifts greatly enhances the overall detectability (see discussion in section \S5).  

\section{Correlation}\label{corr}

\subsection{Angular Clustering of the WHIM}

\begin{figure}
\includegraphics[angle=270,scale=0.4]{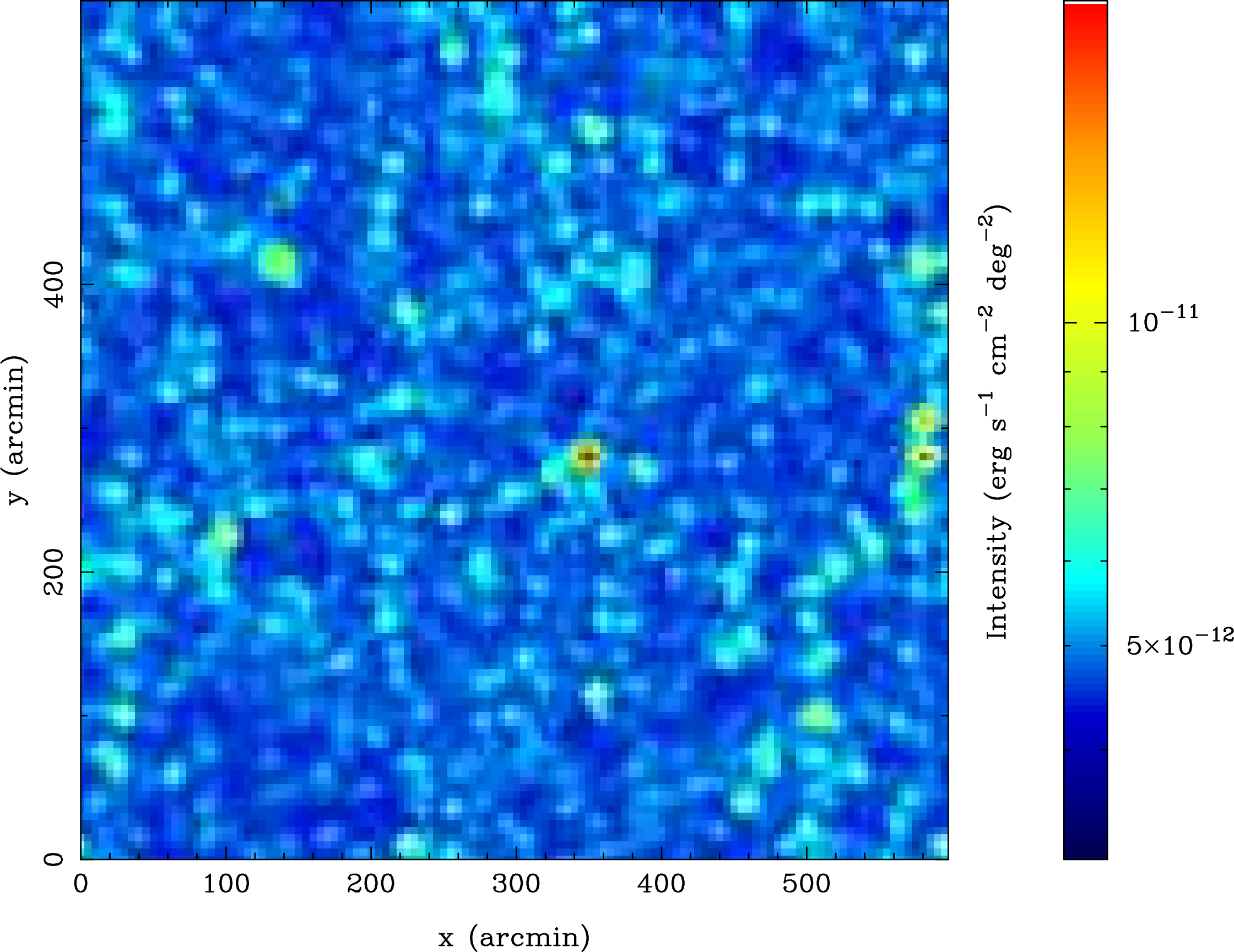}
\caption[h]{A simulation of $10\arcdeg \times 10\arcdeg$ region. The
image is an integration of the total spectrum (IGM plus Galactic
foreground plus AGNs) between 0.5 -- 0.8 keV. The map has been
smoothed with a filter of Gaussian width $\sigma=5\arcmin$. In the
map, emission from many groups of galaxies can be seen as hotspots, as
well as more diffuse emission between them.}
\label{f12}
\end{figure}

As most of the emission from the WHIM comes from relatively low
redshifts, z $<$ 0.3, the strong spatial clustering of the baryons on
small scales is directly reflected in strong angular clustering of
their X-ray emission.  In Figure~\ref{f12} we show a map of a region
of sky of angular size $10\arcdeg \times 10\arcdeg$ taken from our
hydrodynamic simulation. The image is taken in the energy band from
0.5 -- 0.8 keV, and includes an AGN contribution and a galactic
foreground component (both smoothly distributed). The mean flux from
AGN plus foreground is $\sim 4\times 10^{-12}\rm\ ergs\
cm^{-2}s^{-1}deg^{-2}$, which we obtained by integrating the model from
Figure~\ref{f6} between 0.5 -- 0.8 keV. The map has been smoothed
with a filter of Gaussian width $\sigma=5\arcmin$. In the map,
emission from many groups of galaxies can be seen as hotspots, as well
as more diffuse emission between them.

\begin{figure}
\includegraphics[angle=270,scale=0.55]{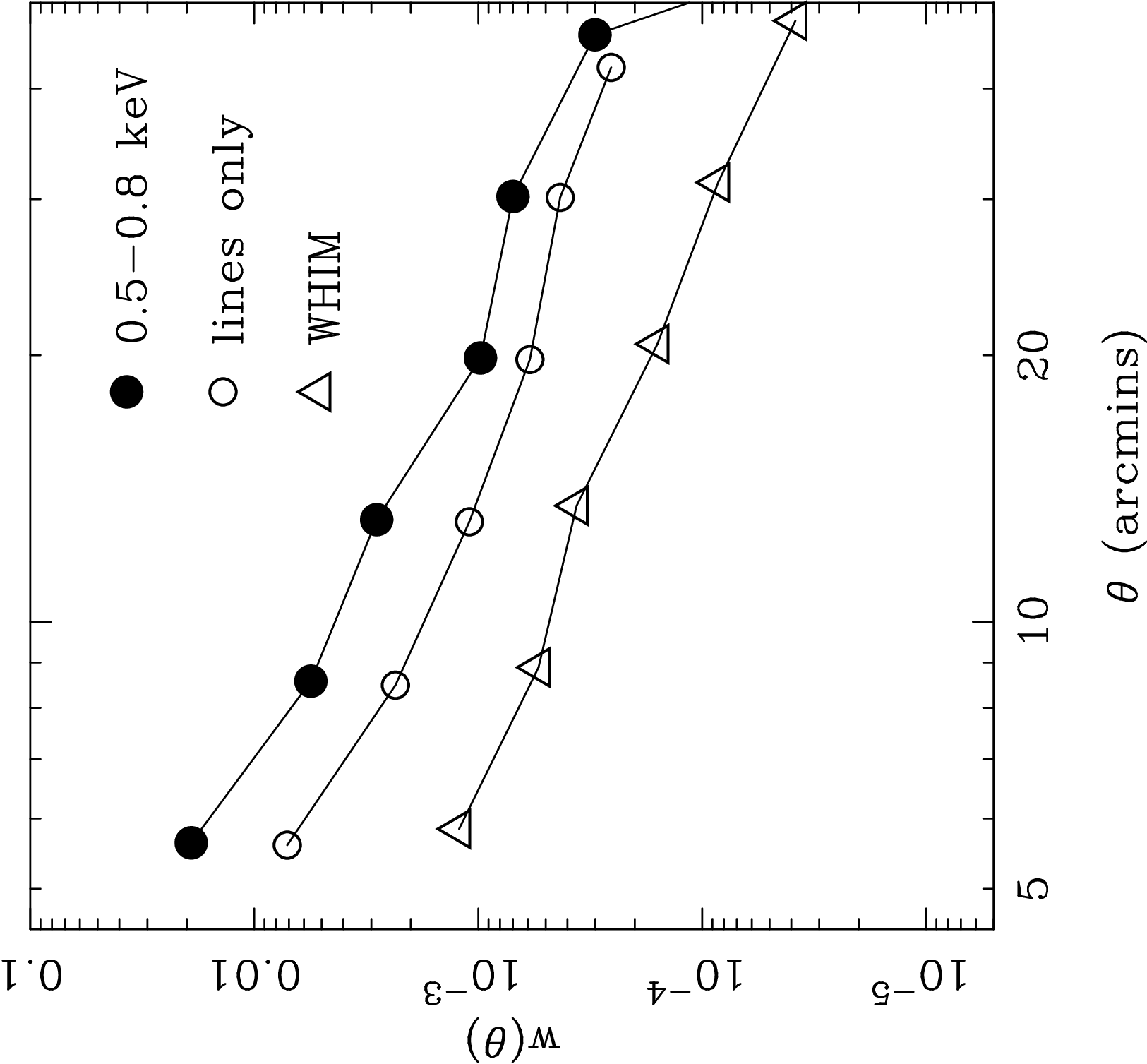}
\caption[h]{Angular auto-correlation function calculated from
Figure~\ref{f12}. X-ray emission has been integrated from 0.5 -- 0.8
keV. Open triangles represent X-ray emission from the WHIM gas ($10^5
< \rm T < 10^7$ K), open circles are for emission lines only, and
filled circles are for X-ray emission from all of the IGM.}
\label{f13}
\end{figure}

The angular clustering of the WHIM emission is treated more
extensively in \citet{cro01}. In this paper we show an
angular map with a larger scale, and in a narrower band.  It is
evident that the WHIM is strongly clustered. We compute the angular
correlation function $w(\theta)$ from this map, where $w(\theta)$ is
defined as:
\begin{equation}
w(\theta) \equiv \left<\delta_X (\alpha) \delta_X
(\alpha+\theta)\right>
\end{equation} Here $\delta_X (\alpha)$ is the X-ray flux over the
mean: $\delta_X \equiv f_X/\left<f_X\right>-1$ where $f_X$ is the
X-ray flux in each pixel between 0.5 -- 0.8 keV. We look at three
types of X-ray emission auto-correlation function: X-rays from WHIM
only, in which we solely use gas particles with
temperatures between $10^5$
-- $10^7$ K; X-rays from the emission lines only, for which we subtract
the continuum from the total spectrum; and X-rays from the IGM. All three types exhibit
a strong clustering signal at high significance (see
Figure~\ref{f13}). This strong clustering is an obvious
signature of the IGM. Photons from AGN originate from much farther
away and their angular clustering is predicted to be smaller on these
scales (see e.g., \citealp{cro01}). Even with only one such field, the
IGM signal shows up very well.

\subsection{Cross-correlation of the IGM Emission and Galaxy Positions}

Numerical simulations show that the X-ray emission from the IGM traces
the same large scale structures as galaxies. By calculating the
angular cross-correlation of the X-ray background (XRB) with nearby
galaxies in simulated surveys of $1\arcdeg\times1\arcdeg$ fields,
\citet{cro01} found a strong signal on an angular separations of
$\theta < 5\arcmin$ (see their Figure~13a). Since the X-ray emissivity
is $\propto n_{e}^{2}T^\alpha$, and the temperature is roughly
proportional to the density for WHIM gas \citep{dav01}, it is obvious
that the large scale pattern of X-ray emission  should correlate well
with that of baryonic overdensity $\delta$.  Moreover, based on the
assumptions in our simulation, the highly ionized metals can be
directly related to local matter overdensities, so that we expect a
strong correlation between emission lines from metals  and nearby
large scale structures. In this section, we will test the idea that
correlation of spectral pixel fluxes with the positions of  galaxies
will enable us to detect the presence of the WHIM.

We first identify galaxies in the simulation using the SKID
groupfinder (see, e.g., \citealp{kwh96})\footnote{see
http://www-hpcc.astro.washington.edu/tools/skid.html}. The galaxies
are then used to create a mock redshift survey occupying the same
simulation space as the X-ray emission. This part of the procedure is
similar to that  employed in \cite{cro01}, and we also produce a flux
limited survey of galaxies. We adjust the magnitude limit so that the
mean redshift of the galaxies is $z=0.2$. We apply an upper redshift
cutoff to  our sample at $z \leq 0.5$, because most X-ray emission
from the hot IGM comes from low redshift. The simulated sky area which
we explore  in this section has a size of
$200\arcmin\times200\arcmin$, and the redshift survey  contains 57,607
galaxies.

We divide the $200\arcmin\times200\arcmin$ field into a $200\times200$
grid, with each cell subtending  $1\arcmin\times1\arcmin$. We
calculate the total spectrum in each cell, accumulating emission from
a redshift up to $z\sim0.5$, and using a spectral resolution of 2
eV.

Our analysis technique involves first subtracting the continuum from
the spectrum to obtain an emission-line only spectrum. We do this
using an iterative technique to lower the emission line profile to the
level of the continuum, without using knowledge of the true
continuum. To study the cross-correlation between galaxies and the X-ray emission lines
from different ion species, for each ion species at a time,  we remap
all the energy bins in the spectrum, which contains emission lines
from all ion species, to redshifts, based on the rest energy of that
particular ion species. For instance, to examine the cross-correlation
between the \ion{O}{8} emission lines and nearby galaxies, we set the
zero-redshift point of each spectrum, which contains emission lines
from all ion species, to $E_0=654$ eV. The redshift $z_{i}$ of
spectral bin with energy $E_{i}$  is assigned according to
$z=E_0/E_i-1$.  Then for each spectral bin with energy $E_i$, we
have a corresponding redshift $z_i$ and two-dimensional angular
coordinate. We then use the redshifts of the spectral bins and their
angular positions on the sky to give each spectral bin a
three-dimensional comoving Cartesian coordinate (making use of the
appropriate relations for  our $\Lambda$CDM cosmology.)  We define the
cross-correlation function $\xi_c(R)$ to be:
\begin{equation}
\xi_c(R) \equiv \left<I_X(r)\delta_g(r+R)\right>.
\label{xi}
\end{equation} Here $I_X$ is the X-ray intensity in each spectral
bin after continuum subtraction, in units of $\rm photons\
cm^{-2}s^{-1}sr^{-1}keV^{-1}$, and $\delta_g$ is the galaxy
overdensity. $\xi_c(R)$ therefore has units $\rm photons\
cm^{-2}s^{-1}sr^{-1}keV^{-1}$.

\begin{figure*}
\includegraphics[width=0.8\textwidth,height=0.7\textheight,angle=90]{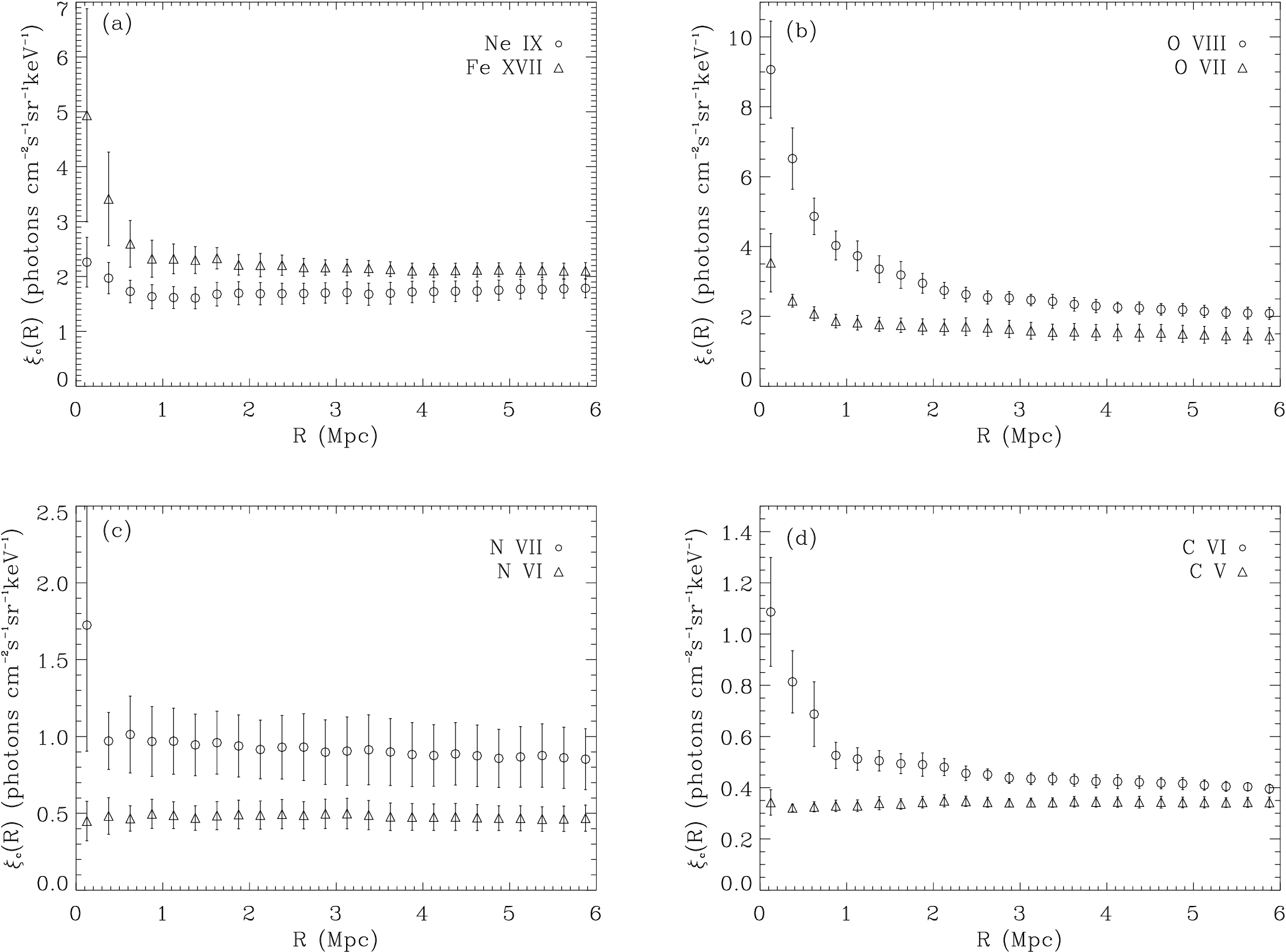}
\caption[h]{Cross-correlation between the IGM emission lines and
galaxies: (a) \ion{Ne}{9} and \ion{Fe}{17}, (b) \ion{O}{7} and
\ion{O}{8}, (c) \ion{N}{6} and \ion{N}{7}, (d) \ion{C}{5} and
\ion{C}{6}. These IGM emission lines (except \ion{N}{6} and
\ion{C}{5}) tend to cluster around galaxies at scales $\lesssim$ 1 --
2 Mpc. }
\label{f14}
\end{figure*}

Figure~\ref{f14} shows the cross-correlation functions for eight
ion species. During the calculation, the zero-redshift points were set
to the corresponding rest-frame energies, as mentioned above. For
He-like ions, we chose the rest resonance energies, and for
\ion{Fe}{17} we let $E_0 = 825.79$ eV. The 1-$\sigma$ error bars are
obtained by splitting the dataset into 20 subsamples, and calculating
the standard deviation of these 20 subsamples. It is apparent that for
all the ions except \ion{C}{5} and \ion{N}{6}, there are strong
signals at small distances, which we interpret to mean that these
X-ray emission lines are strongly correlated with large scale
structures traced by galaxies. In regions close to galaxies, these
signals, after subtracting their mean values at large distances (by
analogy with the galaxy or matter correlation function), can be fitted
with power laws with slopes ranging between -1.2 -- -2.  Since hot
galactic halo gas lies typically within scales less than a few hundred
Kpc from galaxies, we can identify these signals as the correlation
signals of hot gas in the IGM and galaxies.  If we  do not subtract
the mean value of $(\xi_c)$ from the results, the $\log(\xi_c)$ --
$\log(R)$ relation can be approximately  fitted by a power law over
the entire range.  \ion{Fe}{17} and \ion{O}{8} show the steepest power
law slopes of $\sim$ (-0.43), while \ion{O}{7}, \ion{N}{7}, and
\ion{C}{6} show slopes between (-0.33) -- (-0.30). \ion{Ne}{9} has a
rather flat slope of $\sim$ (-0.15).

It is interesting to see that different ions show different
characteristic correlation strengths. The strongest signal comes from
\ion{O}{8} and \ion{Fe}{17}. \ion{O}{7} and \ion{Ne}{9} also show a
relatively strong signal, whereas \ion{C}{6} and \ion{N}{7}  give a
weak signal, while \ion{C}{5} and \ion{N}{6} are not detected at
all. This can be explained by the combination of ion emissivity and
element abundance. From our definition, the correlation function
$\xi_c(R)$ should be a function of $(\epsilon_iZ_i)$, where
$\epsilon_i$ is the line emissivity for ion $X_i$, and $Z_i$ is its
elemental abundance. The higher $(\epsilon_iZ_i)$, the larger the
$\xi_c(R)$. From Figure~\ref{f1}, \ion{O}{8} (solid blue-green
curve) and \ion{Fe}{17} (dotted green curve) show the highest
peak emissivities, so they have the highest correlation
signals. However, since oxygen is more abundant than iron, \ion{O}{8}
gives a stronger signal, even compared to \ion{Fe}{17}. Following a
similar line of argument, we can explain the relative strengths of the
correlation signal from the other ions.

Different ions are detectable over different scales,  depending on the
strength of the emission. The correlation signal for \ion{O}{8} can be
seen strongly for $R \leq 2$ Mpc, while  others (namely \ion{O}{7},
\ion{Fe}{17}, \ion{Ne}{9} and \ion{C}{6}) only start to show an
obvious signal at $R < 1$ Mpc, with \ion{N}{7} only tentatively seen,
at an even smaller scales.  For highly overdense regions, which
typically also have large galaxy overdensities, even on relatively
large scales, the high elemental abundance and large emissivity gives
a significant signal for \ion{O}{8}; while for other ions, the lower
abundances and/or lower emissivities mean that the signal is only
readily seen within short distances from the overdense area, say, $R <
1$ Mpc.

\section{Detectability}\label{dete}

Generally, most emission lines from the filaments are weak, undetected
by current X-ray telescopes. These lines are expected to be embedded
in the strong continuum emission from AGNs plus the Galactic
foreground (see Figure~\ref{f7}). To investigate their
detectability, we define the equivalent width of an emission line at
energy $E$ as
\begin{equation}
EW \equiv \int_{E-\Delta E/2}^{E+\Delta E/2} \left( \frac{I}{I_c} -1
\right) dE,
\end{equation}
where $I$ and $I_c$ are the observed spectrum and its continuum,
respectively, in units of $\rm photons\
cm^{-2}s^{-1}keV^{-1}sr^{-1}$. The integration is over a small energy
range $\Delta E$ around $E$. For a weak emission line, $EW \ll \Delta
E$. Given an exposure time of $T$, the minimum detectable equivalent
width is
\begin{equation}
EW \geq \left(\frac{S}{N}\right) \left(\frac{E}{I_cRA\Omega
  T}\right)^{\frac{1}{2}}.
\label{ew}
\end{equation}
Here $(S/N)$ is the desired signal-to-noise ratio, $R$ is the instrumental resolving power, $A$ is the effective area and
$\Omega$ is the solid angle over which the emission is integrated.

It is apparent, from equation~(\ref{ew}), that an instrument with high
$(RA\Omega)$ is needed to detect weak emission features. Current X-ray
telescopes, such as the {\sl Chandra} X-ray Observatory
\footnote{see http://asc.harvard.edu} and the X-ray Multi-Mirror
Mission ({\sl XMM}-Newton) \footnote{see http://xmm.vilspa.esa.es/},
have unprecedented spectral and spatial resolution. However, the
detectors on board these telescopes either have high spectral
resolution but low effective area, or large effective area but low
resolving power, and thus are inadequate for achieving high resolution
spectroscopy of extended structures that we are interested in.

Several proposed and planned X-ray missions have a large combined
$(RA\Omega)$ and will show promise in detecting weak emission features
from extended structures. We now examine four future X-ray missions,
namely {\sl Astro}-E2 \footnote{see
http://www.isas.ac.jp/e/enterp/missions/astro-eii}, {\sl
Constellation}-X \footnote{see
http://constellation.gsfc.nasa.gov/docs/main.html}, The X-Ray Evolving
Universe Spectrometer (or {\sl XEUS} \footnote{see
http://astro.estec.esa.nl/XEUS}), and the Missing Baryon Explorer (or
{\sl MBE} \footnote{see http://www.ssec.wisc.edu/baryons}). These
X-ray missions are either in the final stage ({\sl Astro}-E2), in planning
stage ({\sl Constellation}-X,{\sl XEUS}), or in proposal stage ({\sl
MBE}). In particular, we will concentrate on {\sl MBE}, a mission
specifically designed to detect diffuse X-ray emission from the WHIM,
to see how this can be achieved.

Table~2 gives the relevant parameters for these four instruments.
In calculating the $EW$, we assume that the desired S/N ratio is 3, that
the line has energy 0.6 keV (where most O and Fe lines lie), and that
the exposure time is 200 ksec. Since the observed spectrum is dominated in
most cases by the emission from the Galactic foreground and the AGN
background, we adopt the continuum value $I_c \approx 24\ \rm photons\
cm^{-2}\ s^{-1}sr^{-1}keV^{-1}$ from Figure~\ref{f6}. We assume that we will devote a time, T,
for each telescope to observe a region of sky of solid angle, $\Omega$, and
we calculate the minimum $EW$ of detectable emission lines integrated over
a solid angle, $\Omega_{PIXEL}$. When a telescope with a field of view,
$\Omega_{FOV}$, maps a region with solid angle, $\Omega$, with a total
exposure time, T, each point of the region receives a exposure time of
$(T\times\Omega_{\rm FOV}/\Omega)$, so the minimum EW in each pixel can be expressed
as:

\begin{equation}
EW \geq \left(\frac{S}{N}\right) \left(\frac{\Omega
  E}{I_cRAT\Omega_{\rm FOV}\Omega_{\rm PIXEL}}\right)^{\frac{1}{2}}.
\label{ew1}
\end{equation}

To compare the minimum detectable $EW$ among the four instruments, we
use as the value for $\Omega_{PIXEL}$ the largest instrument pixel size
($4.9\arcmin \times 4.9\arcmin$ from MBE), and for $\Omega$ we take the largest instrument field
of view ($29.5\arcmin \times 29.5\arcmin$ from MBE). In the case of the same $\Omega$,
$\Omega_{PIXEL}$, and T, equation (5) indicates that the instrument with the
largest value of $RA\Omega_{FOV}$ gives the smallest $EW$, as indicated in the
last two columns of Table~2. Although {\sl XEUS} has much higher angular
resolution than {\sl MBE}, its smaller instrument pixel size offers no
advantage in detecting faint diffuse emission lines, unless the
relevant scale of angular structure is significantly smaller than the
{\sl MBE} pixel size. The large field of view of {\sl MBE} makes it clearly the
best-suited mission for detecting weak X-ray emission lines from
extended sources.

\vbox{ \scriptsize
\begin{center}

\begin{tabular}{lrccrrr}
\multicolumn{6}{c}{Table 2: Instrument Parameters and Detection Limits} \\
  \hline \hline 
Instrument & $A$ & $\Omega_{\rm FOV}$ & $\Omega_{\rm PIXEL} $ & $R$ & $RA\Omega_{\
rm FOV}$ & $EW$ \\  
           & $\rm(cm^2)$ &            &                       &     &
  $(\rm cm^2deg^2)$    & (eV) \\ 
\hline 
{\sl Astro}-E2~$^1$ & 35 &
  $2.9\arcmin\times2.9\arcmin$ & $0.48\arcmin\times0.48\arcmin$ & 100 &
  5 & 138.0 \\  
{\sl Constellation}-X~$^2$ & 3,000 &
  $2.5\arcmin\times2.5\arcmin$ & $5\arcsec\times5\arcsec$ &  400 &
  2,083 & 8.0 \\ 
{\sl XEUS}~$^3$ & 40,000 & $1\arcmin\times1\arcmin$ &
  $2\arcsec\times2\arcsec$ & 500 & 5,556 & 4.9 \\ 
{\sl MBE} & 300 &
  $29.5\arcmin\times29.5\arcmin$ & $4.9\arcmin\times4.9\arcmin$ & 150
  & 10,878 & 3.7 \\  \hline
\label{t2}
\end{tabular}
\parbox{3in}{
\small\baselineskip 9pt
\indent 1. XRS: the X-ray Spectrometer\\ 2. SXT: the Spectroscopy
X-ray Telescope\\ 3. STJ I: the initial configuration for the
Superconducting Tunneling Junctions.  }
\end{center}
\normalsize \centerline{} }

{\sl MBE} is a mission designed specifically to observe the X-ray
emission from the warm-hot, moderate overdense gas predicted to be
distributed in filamentary structures that connect collapsed,
virialized regions such as groups and clusters of galaxies.  It will
also be well suited for detailed studies of emission lines from groups
themselves.  It consists of a single instrument, the X-ray Calorimeter
Telescope (XCT) -- a high-resolution spectrometer and a moderate
resolution imager. The energy resolution is $\sim$ 4 eV (FWHM), with
an energy bandpass from 40 --- 2000 eV.  The field of view consists of
a $6\times6$ array of pixels $\sim\ 4.9\arcmin$ on a side, so that the field of view is
$29.5\arcmin\times29.5\arcmin$.

\begin{figure}
\includegraphics[angle=270,scale=0.4]{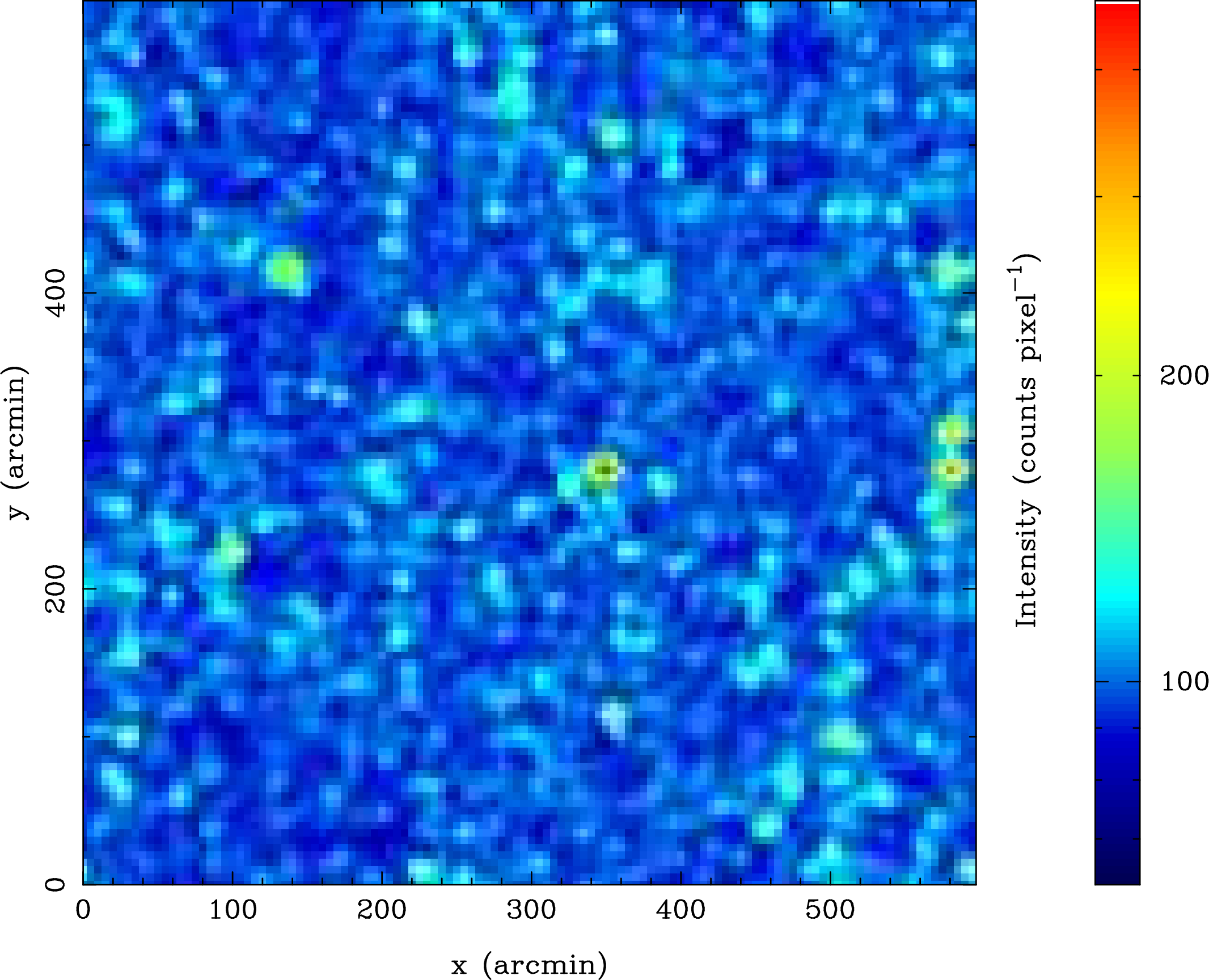}
\caption[h]{{\sl MBE} view of the same area shown in
Figure~\ref{f12}. The total exposure time is $\sim$4 Msec. A
Poisson-distributed noise was added. AGNs plus the Galactic foreground
contribute a mean count of $\sim 80$ counts $\rm pixel^{-1}$}
\label{f15}
\end{figure}

In Figure~\ref{f15} we show a simulated {\sl MBE} count map of the
same $10\arcdeg\times10\arcdeg$ field  shown in Figure~\ref{f12}. The
energy band plotted is between 500 -- 800 eV. The total exposure time
required to build up such an image is $\sim$4 Msec, consisting of 400
pointings of 10 ks each. We add photons from AGNs plus the Galactic
foreground, with a mean count of $\sim 80$ counts $\rm pixel^{-1}$. We
also add Poisson-distributed noise to the map. During such a short
exposure time, the counts collected in each pixel range from $\sim$ 80
-- 90 counts $\rm pixel^{-1}$, for pixels which contain purely
background/foreground emission, up to $\sim 350$ counts $\rm
pixel^{-1}$, arising from  hot spots such as galaxy groups.

\begin{figure*}
\includegraphics[angle=90,scale=0.8]{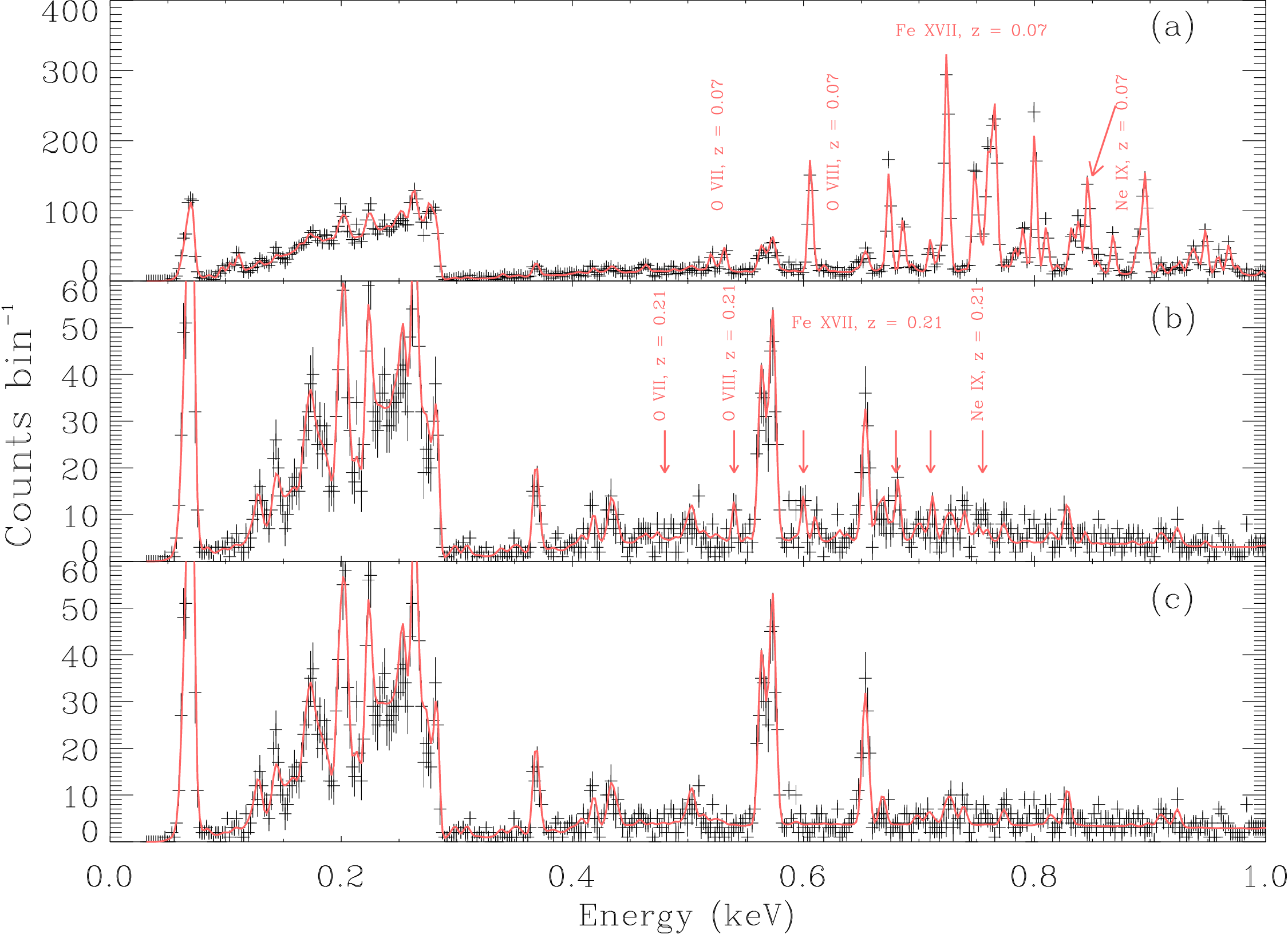}
\caption[h]{Simulated spectra observed with MBE response
functions. The three spectra correspond to the three regions labeled
in Figure~\ref{f3}, respectively. The exposure time is 200 ksec
for each spectrum. In panel (b), it is clear that {\sl MBE} should be
able to probe several weak X-ray emission lines from filaments.}
\label{f16}
\end{figure*}

To demonstrate how {\sl MBE} can detect weak X-ray emission features,
we also present the instrumental-folded spectra of the three areas we
discussed previously (see Figure~\ref{f16}). To obtain these spectra,
we used the software package ISIS (Interactive Spectral Interpretation
System, see \citealp{hde00}) \footnote{see
http://space.mit.edu/ASC/ISIS/}, and applied the instrumental response
data and effective area data  for MBE. The three panels in
Figure~\ref{f16} represent spectra for the three regions we have
examined before. The red lines are the model spectra, and the black
crosses represent observations of these spectra. The errors bars on each point are
based on 1-$\sigma$ Poisson statistics only. The binsize is 2 eV and
the exposure time is 200 ksec. We label the ion species and their
redshifts in red text. It is clear that many weak emission lines,
particularly in Figure~\ref{f16}b, where the emission is from a
filament, can be detected without ambiguity. For instance, the
emission line from the $z\approx 0.21$ \ion{O}{8} line contains nearly
30 counts.

An interesting question to ask is what fraction of the sky as probed
 by {\sl MBE} can yield detectable weak emission lines such as those
 in Figure~\ref{f16}b, given a certain amount of exposure time? This
 we have calculated by examining the 1,600 spectra from the
 $200\arcmin\times200\arcmin$ sky simulation. For most WHIM emission,
 the continuum should be dominated by the AGN plus Galactic foreground
 emission. Taking oxygen lines as an example, assuming most weak
 emission lines seen in Figure~\ref{f7}a span only one bin, which has a binsize of 2 eV, to
 ensure a 4-$\sigma$ detection in one pixel, the signal in the line,
 as a function of exposure time T, should be at least
\begin{equation}
S = 1.85 \times 10^{-2}\ T^{\frac{1}{2}}\ \rm counts,
\end{equation}
and the line peak intensity, also as a function of exposure time T,
should be at least
\begin{equation}
I_{peak} = 2 \times 10^4\ T^{-\frac{1}{2}} + 24\ \rm photons\
cm^{-2}s^{-1}keV^{-1}sr^{-1}.
\end{equation}

\begin{figure}
\includegraphics[angle=90,scale=0.32]{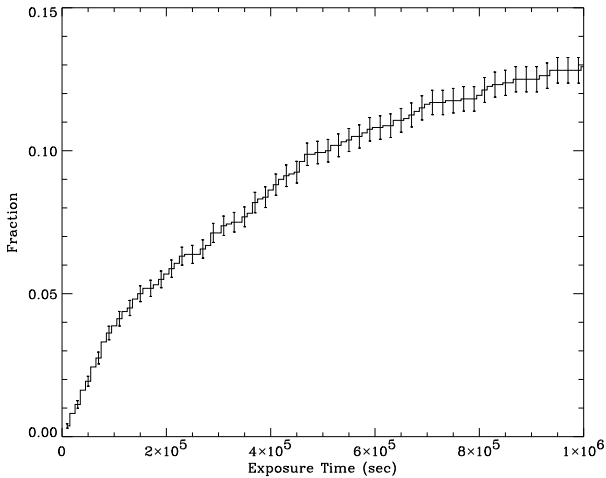}
\caption[h]{The fraction of the sky probed by {\sl MBE} with at least
one detectable (4$\sigma$) weak emission line, given the exposure time
T. The errors are based 1-$\sigma$ poisson fluctuation only.}
\label{f17}
\end{figure}

Assuming an exposure time for each pixel, we can then convert this line peak intensity to line flux sensitivity. Taking $T=200\ ksec$ as an example, the flux sensitivity at  the energy of \ion{O}{8} would then be $1.4\times10^{-10}\rm ergs\,cm^{-2}s^{-1}sr^{-1}$. This is shown as the horizontal dashed line in Figure~\ref{f11}. Typically {\sl MBE} would be able to probe regions with overdensities around 100. In the $200\arcmin\times200\arcmin$ sky simulation, we can then count the total number of angular grid cells for which the spectra show at least one line with peak intensity $\geq I_{peak}$.  The fraction of the sky is obtained by dividing this number by the total number of angular grid cells, which is 1,600 in this case.  Figure~\ref{f17} plots that fraction of the sky as a function of exposure time. The errors on the points are based on 1-$\sigma$ poisson fluctuations only. We can see from this that for one 200 $ksec$ observation  with {\sl MBE}, given the field-of-view of $\sim 900$ square arcmin, roughly $\sim$ 5\% of the field-of-view, or 2 pixels will show detectable emission lines.

\section{Discussion and Summary}\label{disc}

We have presented a detailed study of predictions for the  soft X-ray
emission from the Warm-Hot Intergalactic Medium.  Our main conclusions
can be summarized as follows:

\begin{enumerate}

\item We have examined images of soft X-ray emission from the hot IGM
with a wide field of view ($200\arcmin\times200\arcmin$). Having
spectral information for each pixel allowed us to view the sky in
several different soft X-ray energy bands, where emission lines from
different ion species dominate. The low energy emission is markedly
more diffuse than X-rays in the higher bands.

\item We selected three representative simulated patches of sky to
study with high resolution X-ray spectra: (a) a galaxy group; (b) a
filament; (c) a void-like, underluminous region. In each case, we saw
numerous emission lines from elements such as C, N, O, Fe, Ne, and
identified their redshifts.

\item By taking into account the background X-ray emission from the
AGNs and foreground emission from the Galaxy, we obtained
composite X-ray spectra of the selected regions. We found that in
areas where the galaxy group is present, the X-ray spectrum is
dominated by the hot intragroup medium, while in the void-like areas,
the spectrum is completely dominated by the AGNs plus the Galactic
foreground. In the filament case, although the majority emission comes
from the AGNs plus the Galactic foreground, it is still possible to
detect strong emission lines from the WHIM.

\item We found that the emission in the  (0.5-0.8 keV) band (covering
many of the emission lines)  shows strong small scale angular
clustering, as measured by the angular autocorrelation function. Also,
spectral information for each pixel allowed us to measure three
dimensional clustering. In particular, we computed the
cross-correlation between eight different ion  species and the
simulated galaxy distribution.  We found that \ion{O}{8} showed the
strongest correlation signal, being measurable on scales as large as
$\sim 4-5 h^{-1}$ Mpc, while low energy species such as \ion{C}{5} and
\ion{N}{6} showed no correlation at all. This variation of the correlation
strengths  can be explained by ($\epsilon_iZ_i$), the combination of
line emissivity $\epsilon_i$ and elemental abundance $Z_i$: the larger
($\epsilon_iZ_i$), the stronger the correlation signal.

\item Finally, we studied the detectability of the WHIM with several
proposed X-ray missions. By comparing the effective area, resolving
power and instrumental field-of-view, we found that the Missing Baryon
Explorer ($\sl MBE$) shows the most promise for detecting weak
emission lines from the WHIM gas. We also showed simulated MBE
instrument-folded spectra of our three selected patches of sky.

\end{enumerate}

Accurate prediction of the soft X-ray emission depends on several
important factors, including X-ray emissivity, ionization fraction,
metal abundances, etc. The largest uncertainty in our predictions of
the soft X-ray emission comes from the metal abundance in the
IGM. Theoretically, simulations are beginning to address the formation
and distribution of metals in a self consistent way (see e.g.,
\citealp{cos99,she03}; Aguirre et al. 2001a,b,c). The
metallicity-density relationship we adopted in this paper was
motivated by this work,  and future studies of UV and soft X-ray
emission lines should examine the question of metal enrichment in more
detail (see, e.g., \citealp{fur03}).

Observationally, metals are known to be distributed in a variety of
systems with large dispersion in metallicity $Z$. For example, the
metallicity can be as high as $10Z_{\odot}$ in the central regions of
some active galactic nuclei (AGNs) (\citealp{mdp93,tls97}), and as low
as $10^{-3}Z_{\odot}$ in some halo stars in our galaxy
\citep{bee99}. When we focus on the low redshift IGM, the main
evidence for an enriched IGM comes from the observations of the
X-ray-emitting gas within galaxy clusters
(\citealp{mcc75,mus96,mlo97,dwh00}). With {\sl ASCA}, \citet{mus96}
found that the metallicities in four clusters vary between
$0.3-0.5Z_{\odot}$. Studying $\sim 40$ clusters at $z>0.14$,
\citet{mlo97} concluded that there is no metallicity evolution in
clusters up to $z\sim 0.3$. The observations of the Ly$\alpha$ systems
at low redshifts give metallicities as high as $\sim 0.05Z_{\odot}$
\citep{bty98} indicating a metal enrichment in even lower density
systems at low redshifts. The observations of damped Ly$\alpha$
systems [N(\ion{H}{1}) $> 10^{20}\rm\ cm^{-2}$] show a similar result
\citep{del00}. Due to the lack of observations, the metallicity of the
``WHIM'' is unknown.

Another uncertainty comes from the sensitive dependence of the X-ray
emissivity and ionization fraction on physical properties of the WHIM
gas, such as its temperature and density. For instance, both  the
emissivity and ionization fraction peak over a rather narrow range of 
temperatures, and drop rapidly at low and high
temperatures (see Figure~\ref{f1} and Figure 2, 3, \& 4 of
\citealp{che02}), which means a small change in temperature will cause
order-of-magnitude changes in both quantities. This will in turn cause
a dramatic change in observables such as intensity and emission line
ion species. Because of this the actual observed spectrum depends
sensitively on the underlying physical properties of the WHIM gas.

In this paper, we adopted several approximations to simply the
calculations. First we did not take into account photoionization --
we used collisional ionization when calculating the ionization
fractions. It is important to be aware that photoionization from
background radiation may substantially alter the metal ionization
fractions for low density, low temperature gas (see, e.g.,
\citealp{che02}), which in turn will change the observed
spectrum. However, since most of the emission spectra that can be observed
with {\sl MBE} or other telescopes are produced by hot gas in the high
density, high temperature tail of the WHIM distribution, where
collisional ionization dominates, we expect that photoionization will
not substantially change those observables.

Metal cooling is ignored in our simulation: the cooling is primarily
through H and He. Metal line cooling could be very important in some
cases, especially for high density, warm/hot gas (see, e.g.,
\citealp{sdo93}). However, since our results are very sensitive to the
temperature variation, it is important to quantify the possible effect
of metal cooling on WHIM gas temperature. To cool some hot gas, giving
it a small drop in temperature $\delta T$, the characteristic time
scale $\delta t$ is
\begin{equation}
\delta t = \alpha \frac{k}{n\Lambda (T)} \delta T.
\end{equation}  
Here $\alpha$ is $3/2$ or $5/2$, depending on whether the cooling is
isochoric or isobaric; $k$ is Boltzmann's constant; $n$ is the
electron density; and $\Lambda (T)$ is the cooling function. Assuming
that  we are interested in how cooling acts on timescale of order the
Hubble time, we can estimate the effect of metal cooling on the
temperature decrement $\delta T$ in terms of the physical parameters
of the WHIM gas:
\begin{equation}
\delta \log T \approx 0.1\ n_{-5} \Lambda_{-23} T^{-1}_6 \delta t_{\rm
Hubble},
\end{equation}  
where $n_{-5}=n/(10^{-5}\ \rm cm^{-3})$; $T_6=T/(10^6\ \rm K)$;
$\delta t_{\rm Hubble}$ is the cooling time in units of Hubble time
scale. Here $\Lambda_{-23}$ is cooling rate in units of $10^{-23}\ \rm
ergs\ cm^3s^{-1}$, typical of the WHIM gas with 0.1 solar abundance
\citep{sdo93}. So in general, metal cooling has only a minor effect on
our results; however, since $\Lambda (T) \propto T^{-1}$ in the
temperature range we are interested in, the metal cooling process is
more efficient for low temperature gas, especially so when $\Lambda
(T)$ reaches its peak at $T\approx 2\times 10^5$ K.

It has been known for quite some time that non-gravitational heating,
and/or cooling processes  can change the overall level of cosmological
X-ray emission substantially (\citealp{pen99,wfn01,vbr01a}).  The energy deposited by supernovae in our
simulations has little direct  effect on the IGM. It is added as
thermal energy to particles in high density regions and is radiated
away quickly. Although the direct impact of feedback is therefore
minimal, the cooling included serves to  change the entropy
distribution of gas near the centers of virialized objects and lower
the overall level of emission.  Other treatments of starburst energy
such as those which involve giving wind particles kinetic energy so
that they escape beyond galaxy and group virial radii into the IGM
(e.g., \citealp{she03}) are also likely to change the
distribution of X-ray  emitting gas. Different treatments of feedback
have been shown  by \citet{bvo01} and \citet{vbr01b} to change the pdf of X-ray surface brightnesses in the range which
would affect the WHIM emission lines we focus on here. Varying the
feedback model would be useful in the future in order to assess the
range of effects on the WHIM emission.

Finally, the ionization state of the gas was calculated assuming
ionization equilibrium. Non-equilibrium evolution of the ionization
state can significantly change the ionization fraction (see, e.g.,
\citealp{ech86,hca01}). The characteristic time scale for hot gas to
reach ionization equilibrium, or the recombination time scale is $\sim
(n\alpha_{rec})^{-1}$, where $\alpha_{rec}$ is the recombination
rate. At a temperature of $\sim 10^6$ K, the recombination rate for
\ion{O}{7}, for example, is $\sim 1.4\times 10^{-12}\ \rm cm^3s^{-1}$
\citep{svs82}. Here we include contributions from both radiative and dielectronic recombination. Assuming $n \sim 10^{-5}\ \rm cm^{-3}$, typical of WHIM
gas, the recombination time scale is $\sim$ 2 Gyrs, less than the
Hubble time. However, we must keep in mind that most of the hot gas
was shock-heated at low redshift, so it is very likely that a
significant portion of the WHIM gas has not reached its recombination
time scale and is in a non-equilibrium ionization state.

A reasonable question to ask is: given detected X-ray emission lines,
how do we identify lines with ion species and redshifts? It is
impossible to carry out line identification with a single line;
however, by simultaneously observing multiple lines such as the
\ion{O}{7} triplet and/or \ion{O}{8} redshifted to the same degree,
which is within the capability of the energy resolution of proposed
X-ray missions such as {\sl MBE}, we can easily identify lines with
redshifts. In addition, measurements of the line ratios can provide
direct temperature/density diagnostics for the ionized plasma (see,
e.g., \citealp{vcm86}). Observations of soft X-ray emission lines have
the potential to not only reveal the presence of the missing baryons
but to give us detailed information on their clustering, temperature,
and  metallicity.
 
X-ray emission provides a direct probe of the WHIM gas. Unlike the
X-ray absorption method, which can yield only one dimensional
information, by combining imaging/spectroscopic observations, X-ray
emission can show us the fully three-dimension distribution of the
WHIM gas. This is essential not only for closure of the local cosmic
baryon budget, but also if we are to understand the distribution of
the WHIM gas, and how structures form and evolve. Moreover, by
applying line diagnostic methods, X-ray emission lines can be used to
probe the metallicity, temperature and density of X-ray emitting
gas. This in turn can give us important information on galaxy
formation and evolution, given that most of these metals which are
responsible for X-ray emission lines are produced by supernova
explosions and propagate from the interstellar medium into
intergalactic medium. Although the relatively low density and
temperature make it hard to observe, currently proposed X-ray missions
will for the first time provide us with detectable signals from the
X-ray emission of the WHIM gas.

This work is supported by Carnegie Mellon University. T.F. thanks members of the astrophysics group at CMU for their help and stimulated discussions. T.F. also thanks to the hospitality of the Kavli Institute for the Theoretical Physics at the University of California, Santa Barbara, where part of the work was conducted. This research was supported in part by the National Science Foundation under Grant No. PHY99-0794.

{}

\end{document}